\documentclass[onefignum,subeqn,final]{siamltex}
\usepackage{amsfonts}\usepackage{epsfig}
\psfull
\title{%Mathematical model for living media with hierarchically
Cooperative mechanism of self-regulation in hierarchical living systems%
\thanks{This work
was made possible in part by
Grants U1I000, U1I200 from the International Science Foundation,
Grant 96-02-17576 from the Russion Foundation of Basic Researches and
Grant 4.4/103 of Fundation of Fundamental Researches of Ukraine.}}

\author{I.~A.~Lubashevsky%
\thanks{Department of Physics, Moscow State
University, Vavilova str., 46--92, 117333 Moscow, Russia}
\and V.~V.~Gafiychuk%
\thanks{Institute for Applied Problems
of Mechanics and Mathematics,
National Academy of Sciences of Ukraine,
3b Naukova str. L'viv, 290601, Ukraine}
}

\begin{document}
\maketitle

\begin{abstract}

We study the problem of how a ``living'' system complex in structure
can respond perfectly to local changes in the environment. Such a
system is assumed to consist of a distributed ``living'' medium and a
hierarchical ``supplying'' network that provides this medium with
``nutritious'' products. Because of the hierarchical organization
each element of the supplying network has to behave in a
self-consistent way for the system can adapt to changes in the
environment.

We propose a cooperative mechanism of self-regulation by which the
system as a whole can react perfectly. This mechanism is based on an
individual response of each element to the corresponding small piece
of the information on the state of the ``living'' medium. The
conservation of flux through the supplying network gives rise to a
certain processing of information and the self-consistent behavior of
the elements, leading to the perfect self-regulation. The
corresponding equations governing the ``living'' medium state are
obtained.

\end{abstract}



\pagestyle{myheadings}
\thispagestyle{plain}
\markboth{I.~A.~LUBASHEVSKY AND V.~V.~GAFIYCHUK}
{COOPERATIVE MECHANISM OF SELF-REGULATION}

\section{Introduction. ``Living'' systems and the self-regulation problem}

The present paper is devoted to one of the fundamental problems of
how complex ``living'' systems widely met in nature can adapt to
changes in the environment. By the term ``living'' system we mean one
that comes into being, provides for itself, and develops pursuing its
own goals. This class comprises a great variety of biological and
ecological systems. Besides, economic systems can be also placed into
such a class because their occurrence, growth, and development is due
to self-organization processes. To make the subject of our analysis
more clear let us, first, specify the main properties of the systems
we deal with.

\subsection{Living systems}

The elements of such a system, first, should be permanently supplied
with external ``nutritious'' products for their life activities.
Second, in order for its elements to subsist special conditions are
required.  The latter are implemented through these elements making
up a certain medium (which will be referred below as to a ``living''
medium) whose state is controlled by keeping its basic parameters
inside certain vital intervals.

As a rule both of these requirements are fulfilled by a supplying
network which provides the elements with nutrients as well as
controls the living medium state. Since the external products usually
penetrate into the system through a common entryway and, then, should
be delivered to a great number of elements the flow of these products
has to branch many times until it reaches these elements. So the
supplying network is to be organized hierarchically and to involve
many levels.

Let us, now, present typical examples of such systems. First, this is
living tissue, where blood flowing through a vascular network
supplies cells with oxygen, nutritious products, etc. At the same
time blood withdraws carbon dioxide and other products resulting from
the cell life activities, keeping their concentrations inside the
vital intervals (see, e.g.~\cite{2}). In this way blood flow controls
also the tissue temperature \cite{1}. We note that the temperature
and the concentration of carbon dioxide (or a similar substance) are
major parameters characterizing the living tissue state \cite{2},
because their values are directly determined by the life activities
of cells. In order to supply with blood each small group of cells the
vascular network or, more precisely, its arterial and venous beds are
approximately of the tree form and involve many vessels of various
lengths and diameters. The regional self-regulation processes keep
the carbon dioxide concentration and the temperature inside certain
vital intervals only in which living tissue can function normally
(see, e.g.~\cite{2,1}).

A similar example is a respiratory system where oxygen going through
a hierarchical network of bronchial tubes reaches small capillaries.

Second, large firms are a clear example of economic hierarchical
systems.  Managers of all functions and of all levels make up a
management network \cite {3}. Roughly speaking, the management
network controls both the money flow toward the organization low
level comprising workers and also the flow of products in the
opposite direction. In performing technological processes the wages
paid to workers actually transform into the firm products.

The existence of tremendous amount of goods in the market, in
contrast to a relatively small number of raw materials shows that
there also must be large hierarchical systems supplying the consumers
with goods. The goods flow after reaching the consumers transforms
into money flow in the opposite direction \cite{3a,3b}. In
particular, self-organization phenomena can give rise to trade
networks of the tree form \cite{Posp}.

Concerning ecological systems we would like to note that they are
also complex in structure and can involve a larger number of
``predator-prey'' levels \cite{4}. The dynamics of ecosystems is
governed by biomass and energy flow on trophic networks and under
certain conditions leads to formation of the tree like trophic
networks \cite{PhRev}.

\begin{figure}[t]
\begin{center}
\epsfig{file=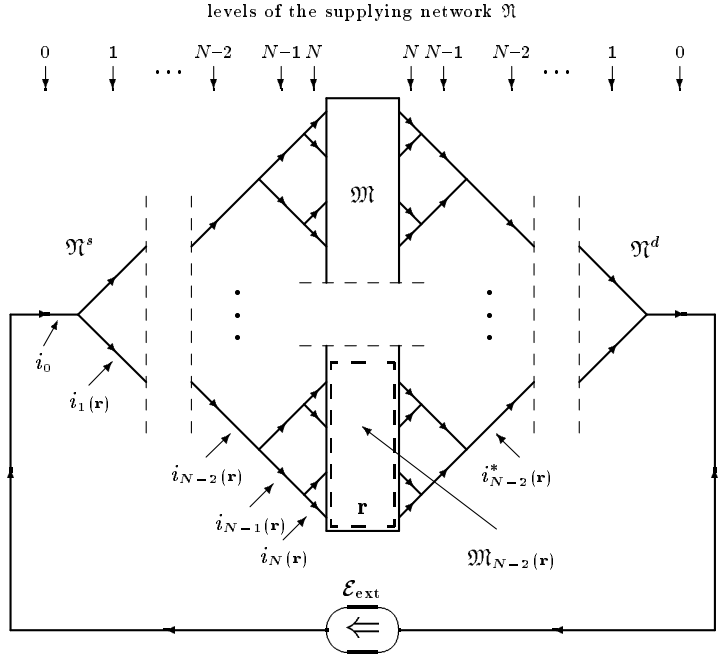}
%\small \input fig1.pic
\end{center} \caption{``Living'' system structure.
$($For simplicity is shown the binary network.$)$}
\label{F1}
\end{figure}

Keeping in mind the aforementioned examples let us represent a living
system as one consisting of a living medium and a supplying network which
involves supplying and draining beds of the tree form (Fig.~\ref{F1}). (In
principle, the two beds can coincide with each other in space.) A transport
agent (e.g., blood in living tissue) ``flowing'' through this network
delivers ``nutrients'' to the living medium and withdraws ``life activity''
products.  The interaction between the transport agent and the living
medium (leading to the product exchange) takes place when the transport
agent ``flows'' through the last level branches of the supplying network
(in Fig.~\ref{F1} through the branches of level $N$\/).

The concentration of the life activity products will be regarded as the
parameter characterizing the living medium state. This is due to the fact
that the higher are life activities in intensity, the greater is the
concentration of the resulting products. So an increase in this
concentration informs the system that the living medium needs a greater
amount of ``nutrients''. In addition, the products resulting from the life
activities and remaining in the living medium themselves can depress its
functions. Therefore the system has to prevent the concentration of these
products exceeding a certain critical value.

The motion of transport agent is accompanied by energy dissipation.
So a certain external force should be applied to the system that
affects the overall flow of transport agent (for example, in living
tissue this is the blood pressure \cite{2,1}). Further distribution
of the transport agent flow over the supplying network is governed by
the individual properties of its branches. More precisely, the
transport agent flow distribution is controlled by both a certain
potential (in living tissue it is the blood pressure) and by the
``resistances'' of the branches to the transport agent flow. In order
to specify the physical regularities of the transport agent flow
distribution we may choose various approaches. One of them is
Prigogine's principle of minimum entropy production stated in
nonequilibrium thermodynamics (see, e.g., \cite{7}). Another is
related to Swenson's principle of maximal dissipation. In
our analysis the two approaches lead to the same result: the
transport agent flows meet an extremum of a certain functional. So,
to be defined we will adopt the minimum entropy production principle.

\subsection{Self-regulation problem}

For different elements of the living medium to fulfill their
individual functions independently of one another the supplying
network should, at least in the ideal case, have a capacity for
controlling the state of the living medium at each its point. This
requirement is reduced to the local control of the ``perfusion'' rate
of the transport agent through the living medium. So the independence
of the perfusion rate at one point from life activities of elements
located at other points is a desirable property which will be
referred below as to the perfect self-regulation. In fact, if we
ignore diffusion in the living medium solely the perfusion rate of
transport agent at a given point will govern functioning of elements
at this point: supply them with ``nutrients'' and withdraw the life
activity products. So the perfusion rate should take such a value
that provides the optimal conditions for functioning of the elements,
depending on their life activities.  Therefore if the perfusion rate
at one point reacted substantially to the life activities at other
points then the living medium elements would interfere with one
another and the living system could lose its capacity for adapting.
The existence of a great variety of living systems in nature enables
us to think that such a nonlocal interaction is suppressed or, at
least, depressed to a certain degree. However, in trying to describe
how this self-regulation can be implemented, i.e. its particular
mechanism we meet the following fundamental problem.

The perfusion rate of the transport agent at a given point ${\mathbf
r}$ is certain to be determined by its flow through the corresponding
last level branch $ i_N({\mathbf r})$ of the supplying network
(Fig.~\ref{F1}). However, this flow in turn depends on the transport
agent flow through the branch $i_{N-1}({\mathbf r})$ of the previous
level connected with the branch $i_N({\mathbf r})$ and so on up to
the stem $i_0$ (the branch of zeroth level). In this way we find a
path ${\mathbb P}({\mathbf r})=\{i_0,i_1({\mathbf r}),\ldots
,i_{N-1}({\mathbf r}),i_N({\mathbf r })\}$ on the supplying network
through which the transport agent flow reaches a small neighborhood
of the point ${\mathbf r}$. Let us consider the transport agent flow
through one of these branches, for example, a branch $ i_n({\mathbf
r})$ of level $n<N$ (in Fig.~\ref{F1} it is the branch $i_{N-2}(
{\mathbf r})$). This branch supplies with ``nutrients'' not only
elements in the vicinity of the point ${\mathbf r}$ but also all of
the elements located inside a certain domain ${\mathfrak
M}_n({\mathbf r})$ (inside the domain ${\mathfrak M}_{N-1}({\mathbf
r})$ in Fig.~\ref{F1}) and controls the living medium state in this
domain as a whole. So the required flow of the transport agent through
the branch $i_n( {\mathbf r})$ is specified by the life activities of
all the elements belonging to the domain ${\mathfrak M}_n({\mathbf
r})$ and a change in the life activities of one of them inevitably
will cause this flow to vary.

Since the external force applied to the system controls only the
overall flow the specific distribution of the transport agent flow
over the supplying network is governed by the ``resistances'' of all
the branches.  Under such conditions, in general, a change in the
transport agent flow through the branch $i_n({\mathbf r})$ (for
example, because of variations in its ``resistance'') causes the
transport agent flow to alter in all the branches of higher levels
that are connected with $i_n({\mathbf r})$, in particular, in a last
level branch $i_N({\mathbf r}^{\prime })$ leading to a point
${\mathbf r} ^{\prime }\in {\mathfrak M}_n({\mathbf r})$ and
${\mathbf r}^{\prime }\neq {\mathbf r}$. So, in principle, a change
in the life activities of elements near the point $ {\mathbf r}$ will
lead via the branch $i_n({\mathbf r})$ to variations in the perfusion
rate of transport agent at all the points belonging to the domain $
{\mathfrak M}_n({\mathbf r})$. Since the domains $\left\{ {\mathfrak
M}_n({\mathbf r})\right\} $ increase in size and tend to the total
space of the living medium as we pass on the supplying bed from the
last level to its stem this effect is nonlocal substantially.

Therefore special conditions are required for the self-regulation to
be perfect because in the general case the perfusion rate of the
transport agent is determined by the life activities of all the
elements.

In other words for the perfect self-regulation to occur the
``resistances'' of all the branches should vary in a self-consistent
way. In this place we actually meet the fundament problem mentioned
above: what governs such variations of the branch ``resistances''?
For a large living system it is unlikely that any one of its elements
can possess the whole information required of governing the perfect
response to changes in the environment and, so, in the life
activities of living medium. This is due to the fact that such a
control requires processing of a great amount of information
characterizing the living medium state on all spatial scales (i.e. at
all the levels of supplying network). In particular, in living
organisms local alterations in homeostasis seem to be controlled by
regional mechanisms of self-regulation rather than the central
nervous system \cite{2}. Besides, none of these elements can
individually control transport agent flow through the supplying
network because of the mass conservation at its branching points.

The aforementioned allows us to assume that the perfect
self-regulation can be implemented through cooperative mechanisms. By
this scenario each branch of the supplying network receives the
corresponding small piece of the information on the living medium
state and their reaction to this information gives rise to the
desired redistribution of the transport agent flow over the supplying
network. In other words, the branch ``resistances'' vary in such a
self-consistent way that enables the supplying network to provide,
for example, an additional amount of ``nutrients'' for the living
medium elements which have a need for this and not to disturb
transport agent flow at other points of the living medium.

However, for the cooperative mechanism of perfect self-regulation to
come into being, first, a certain self-processing of the information
is required which enables the branches to react adequately, depending
on their place in the supplying network. Second, the physical
properties of the supplying network should give rise to the
cooperative effect of variations in the branch ``resistances'' on the
transport agent flow redistribution leading to the desired results.

The purpose of the present paper is to show that these requirements
can be fulfilled and so a cooperative mechanism of perfect
self-regulation in living systems can exist. We will demonstrate that
there is a required self-processing of information and the physics of
transport agent motion ensures its proper redistribution over the
supplying network. Beforehand we can say that it is the mass
conservation at the nodes of supplying network and its hierarchical
organization those bring into being the information self-processing.
This information self-processing is implemented through measuring the
concentration of the life activity products inside each branch of the
draining bed. Due to the interaction with the living medium the
transport agent saturates with the life activity products when it
flows through the last level branches. Then the transport agent moves
through the draining bed from higher hierarchy levels to lower ones
and, so, at the nodes its smaller streams unite into larger streams.
Therefore, due to the mass conservation the concentration $\theta _i$
of life activity products inside, for example, a branch $i$ is
actually equal to the concentration of life activity products inside
the living medium averaged over the domain $ {\mathfrak M}_i$ which is
drained as a whole by this branch. So the value of $\theta _i$
aggregates the information on the living medium state in the domain $
{\mathfrak M}_i$ as a whole and, thus, can play the role of the information
piece needed for this branch and the corresponding branch of the
supplying bed to react properly to variations in the life activities
of the living medium. The correspondence between the branches of the
supplying and draining beds is illustratied in Fig.~\ref{F1} by the pair
$i_{N-2}(\mathbf r)$, $i^*_{N-2}(\mathbf r)$.

It should be noted that developing this model we actually have kept
in mind one of the possible mechanisms of self-regulation in living
tissues. Life activities of cellular tissue gives rise to variations
in the carbon dioxide concentration in it and, so, in venous blood.
Receptors embedded into the vein walls through a regional nervous
system govern the expansion or contraction of the corresponding
arteries (see, e.g., \cite{2,1}).

Below, at first, we will analyze in detail the model for a living
system with a regular supplying network. This network is organized in
such a manner that the perfusion rate of transport agent be the same
at every point of the living medium, all other factors being equal.
In particular, we will develop a technique that not only enables us
to study the perfect response of living system to changes in the
living medium state but also can be used in investigations of its
more complex behavior. Then we will show that the perfect
self-regulation can also occur in living systems with nonregular
supplying networks of the general form.

\section{Mathematical model for living system with regular supplying
network\protect\label{sec:model}}

Let us consider a system consisting of a living medium ${\mathfrak M}$ and
a hierarchical supplying network ${\mathfrak N}$ similar to that shown in
Fig.~\ref{F1}. The network ${\mathfrak N}$ involves supplying $({\mathfrak
N}^s)$ and draining $( {\mathfrak N}^d)$ beds of the tree form (illustrated
in Fig.~\ref{F1} by the left and right--hand side parts) and an
external element ${\mathcal E}_{\mathrm{ext}}$ joining the tree stems.
Transport agent flowing through the former bed supplies the living
medium ${\mathfrak M}$ with ``nutrients''. At the same time transport agent
withdraws life activity products from the living medium through the
draining bed. For simplicity we assume that the living medium domain
${\mathfrak M}$ is a $d$-dimensional cube of edge $l_0$ and the geometry
of both the beds is the same.

The supplying and draining beds can be represented as the collections
of groups of branches $\{i\}$ belonging to one level: ${\mathfrak
N}^s=\{{\mathbb L} _n^s\}_{n=0}^N$, ${\mathfrak N}^d=\{{\mathbb
L}_n^d\}_{n=0}^N$, where ${\mathbb L} _n^{s(d)}=\left\{ i:i\in {\mathfrak
N}^{s(d)}\ \mathrm{and}\ i\in \mathrm{level}~n\right\} $.
Below we will use the symbol ${\mathbb L}_n$ instead of
${\mathbb L}_n^s$ or ${\mathbb L}_n^d$ where it does not lead to
misunderstanding or gives the same results due to the mirror symmetry of
the supplying and draining beds. In these terms the embedding of the
supplying network $ {\mathfrak N}$ (as the collection of branches $\{i\}$)
into the living medium matches the partition of the domain ${\mathfrak M}$
into the following collection of cubes $\{\{{\mathfrak M}^i\}_{i\in
{\mathbb L}_n}\}_{n=0}^N$.

For each level $n$ the group ${\mathbb M}_n=\{{\mathfrak M}^i\}_{i\in
{\mathbb L}_n}$ involves $2^{nd}$ equal disjoint cubes of edge
$l_n=l_02^{-n}$ that together compose the whole living medium domain:

\begin{equation}
\label{em1}{\mathfrak M}=\bigcup_{i\in {\mathbb L}_n}{\mathfrak M}^i\quad
\mathrm{and}\quad {\mathfrak M}^i\bigcap {\mathfrak M}^j=\emptyset \quad
\mathrm{if}\quad i\neq j,\quad i,j\in {\mathbb L}_n.
\end{equation}
For any two levels $n$ and $m<n$ each domain ${\mathfrak M}^i\in {\mathbb
M}_n$ belongs to one of the domains of the group ${\mathbb M}_m$ and,
thus, each domain ${\mathfrak M}^j\in {\mathbb M}_m$ can be represented as a
certain union of the domains of group ${\mathbb M}_n$

\begin{equation}
\label{em2}{\mathfrak M}^j=\bigcup_{i\in {\mathbb L}_n^j}{\mathfrak M}^i.
\end{equation}
For zeroth level $\mathfrak M^{i_0}=\mathfrak M$. Below the cubes
of the group ${\mathbb M}_n=\{{\mathfrak M}^i\}_{i\in {\mathbb L}_n}$ will be
also called fundamental domains of level $n$ and those of the last
level $N$ will be also referred to as elementary domains.

The supplying and draining beds are embedded into the living medium
in such a way that every branch $i$ of a given level $n$ supplies (or
drains) a certain domain ${\mathfrak M}^i\in {\mathbb M}_n$ as a whole. Each
elementary domain is bound up with one of the last level branches.
The last level number $N$ is assumed to be much larger than unity:
$N\gg 1$, and the length $l_N$ may be regarded as an infinitely small
spatial scale. Transport agent flow is directed from lower to higher
levels on the supplying bed and in the opposite direction on the
draining bed.

\begin{figure}[t]
\begin{center}
\epsfig{file=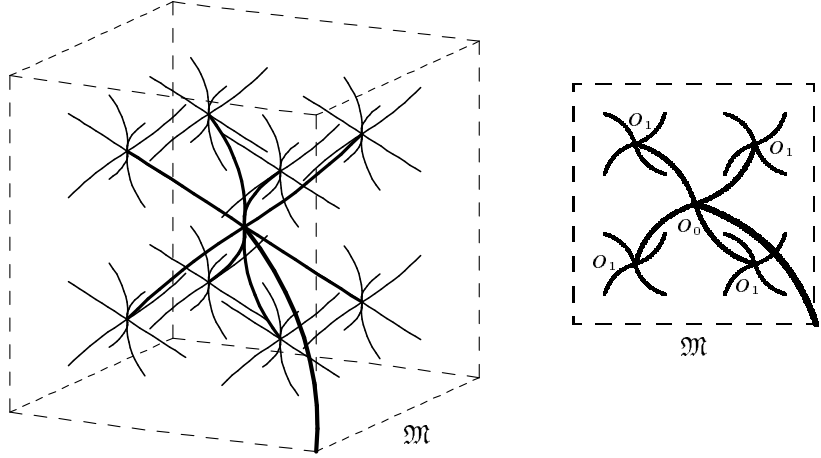}
\end{center}
\caption{Fragments of the supplying network architectonics for the
two- and three-dimensional living medium}
\label{F2}
\end{figure}

Since the geometry of both the beds is assumed to be the same, we
specify it for the supplying bed only. The stem of this bed
(Fig.~\ref{F2}) goes into the cube ${\mathfrak M}$ through one of its
corners and reaches the cube center $ O_0$, where it splits into
$g=2^d$ branches of the first level. Each branch of the first level
reaches a center $O_1$ of one of the $g$ fundamental domains of the
first level. At the centers $\{O_1\}$ each of the first level
branches in turn splits into $g$ second level branches. Then this
process is continued in a similar way up to level $N$. The branches
of the last level $ N $ are directly connected with living medium
${\mathfrak M}$. It should be noted that in this model the length $\ell _n$
of branches belonging to level $n$ is estimated by the expression
$\ell _n\sim \sqrt{d}l_n=\sqrt{d}2^{-n}$ and smaller is a branch, to
a higher level belongs it.

Transport agent does not interact with the living medium ${\mathfrak M}$
during its motion through the branches except for the last level
ones. When the transport agent reaches one of the last level branches
of the supplying bed, for example, branch $i_N$ the ``nutritious''
products carried with it uniformly spread over the elementary domain
${\mathfrak M}_N^i$ containing the branch $i_N$. At the same time transport
agent is saturated with the life activity products, withdrawing them
from the domain ${\mathfrak M}_N^i$. In this way the life activity products
going with the transport agent through the draining bed leave the
system.

The state of the medium ${\mathfrak M}$ is described by a certain field
$\theta ( {\mathbf r},t)$ being the dimensionless concentration of life
activity products.  Dynamics of the field $\theta ({\mathbf r},t)$ is
governed by the volumetric generation $q({\mathbf r},t)$ resulting from
the life activities, the dissipation because of draining the living
medium, and the diffusion. In other words the field
$\theta (\mathbf r, t)$ is
considered to evolve according to the equation

\begin{equation}
\label{*2.1}\frac{\partial \theta }{\partial t}=D\nabla ^2\theta +q-\theta
\eta ,
\end{equation}
where $D$ is the diffusivity and $\eta ({\mathbf r},t)$ is the perfusion
rate of transport agent. Since a branch $i_N$ of the last level
supplies the elementary domain ${\mathfrak M}_N^i$ of the living medium as
a whole the transport agent flow $J_{i_N}$ through the branch $i_N$
and the perfusion rate $\eta ({\mathbf r},t)$ are related by the
expression

\begin{equation}
\label{*2.2}J_{i_N}=\int\limits_{{\mathfrak M}_N^i}\eta (\mathbf r)
d{\mathbf r.}
\end{equation}
The distribution of the transport agent flow $\{J_i\}$ over the
network $ {\mathfrak N}$ obeys the mass conservation at its nodes
$\{{\mathfrak b}\}$

\begin{equation}
\label{*2.4}(J_i)_{\mathrm{in}}=\sum\limits_{i\in {\mathfrak b}}
^{\mathrm{out}}J_i \qquad \mathrm{or}\qquad
\sum\limits_{i\in{\mathfrak b}}^{\mathrm{in}} J_i= (J_i)_{\mathrm{out}},
\end{equation}
where $J_i$ is the
transport agent flow through a branch $i$ going in or out of the node
${\mathfrak b}$  and the sums run over all the branches $i\in
{\mathfrak b}$ leading from or to this node for the supplying and
draining beds, respectively.  Provided the field $\eta ({\mathbf r},t)$
is given expressions~(\ref{*2.2}) and (\ref{*2.4}) completely specify
the quantities $\{J_i\}$.

The last term in equation (\ref{*2.1}) implies that the transport
agent when going through the branch $i_N$ is saturated with the life
activity products up to their concentration in the elementary domain
${\mathfrak M}_N^i$. Then these products moves with the transport agent
through the draining bed without exchange with the living
medium. Therefore, we also ascribe to the transport agent going
through the draining bed the collection of variables $ \{\theta _i\}$
measuring the concentration of the life activity products inside the
branches $\{i\in {\mathfrak N}^d\}$ and regard the value $J_i\theta _i$ as
the flow of these products in the branch $i$. For a branch $i_N$ of
the last level we write

\begin{equation}
\label{*2.3}J_{i_N}\theta _{i_N}=\int\limits_{{\mathfrak M}_N^i}
\theta (\mathbf r)
\eta (\mathbf r) d{\mathbf r}.
\end{equation}
and assuming the conservation of the life activity products at the
nodes of the draining bed we get

\begin{equation}
\label{*2.5}\sum\limits_{i\in {\mathfrak b}}^{\mathrm{in}}
J_i\theta _i=(J_i\theta_i)_{\mathrm{out}}.
\end{equation}

Equation (\ref{*2.1}) among with expressions (\ref{*2.2}) and
(\ref{*2.3}) describe the dynamics of the state of the
the living medium and its
interaction with transport agent. Equations (\ref{*2.4}) and
(\ref{*2.5}) reflect the general laws of transport phenomena in the
supplying and draining beds.

Let us, now, specify the regularities governing the transport agent
flow through the network ${\mathfrak N}$. First, provided the
characteristic parameters of branches are fixed the transport agent
flow patter $\{J_i\}$ is assumed to meet the minimum condition of
the total energy dissipation due to transport agent motion:

\begin{equation}
\label{min}{\mathbb D}\{J_i\} \Rightarrow \min_{\{J_i\}}
\end{equation}
subject to relationships (\ref{*2.4}), where the total rate of energy
dissipation is given by the expression

\begin{equation}
\label{*2.6}{\mathbb D}\{J_i\}=\frac 12\sum\limits_iR_iJ_i^2-
J_0{\mathcal E}_{ \mathrm{ext}}.
\end{equation}
Here $R_i$ is the kinetic coefficient characterizing
the energy dissipation during the motion of transport agent through
the branch $i$, ${\mathcal E}_{ \mathrm{ext}}$ is the external force
causing the transport agent motion through the network ${\mathfrak N}$ as a
whole, $J_0$ is the transport agent flow through the tree stems, and
the sum runs over all the branches of the network ${\mathfrak N}$. It
should be noted that previously by the term ``branch resistances'' we have
exactly meant the given kinetic coefficients because they do play the
role of the true branch resistances to transport agent flow as it
will be shown below in the next section.

Second, in this model it is the coefficients $\{R_i\}$ those
characterize the individual effect of the branches on the transport
agent flow. So the system response to changes in the living medium
state, namely, to variations in the field $\theta ({\mathbf r},t)$ (see
Introduction) is represented as time variations of the
coefficients $\{R_i\}$. Since the supplying and draining beds have
been assumed to be equivalent in architectonics the flow pattern on
these beds $\{J_i\}$ as well as the coefficient collection $\{R_i\}$ are
set to be the mirror images of each other within reversing the flow
direction.  The latter allows us to confine our description of the system
response to the draining bed.

As mentioned in Introduction the pattern $\{\theta _i\}$ can be
treated as the aggregated information on the living medium state on
all the spatial scales. In particular, the variable $\theta _i$
corresponding to the branch $ i$ characterizes the state of the
living medium in the fundamental domain $ {\mathfrak M}^i$ as a whole.
Therefore we assume that for each branch, for example, a branch $i$
of level $n$ time variations in the coefficient $R_i$ are directly
controlled by the variable $\theta _i$ assigned to this branch.
Under the steady-state conditions this means that the coefficient
$R_i$ is an explicit function of the variable $\theta _i$ identical
for all the branches of one level $n$, i.e. $R_i=R_n(\theta _i)$.
The general properties of the $R_n(\theta )$ dependence are
actually determined by the fact that the transport agent flow through
the network ${\mathfrak N}$ should grow as the variables $\{\theta _i\}$
increase. Indeed, let changes in the environment cause the life
activities of the living medium to grow in intensity. The latter
leads immediately to an increase in the concentration $\theta ({\mathbf
r},t)$ of life activity products inside the living medium, causing an
increase in the variables $\{\theta _i\}$. The higher is the life
activity intensity, the greater is the amount of ``nutrients'' needed
for the living medium.  Therefore, under such conditions the
transport agent flow through the network ${\mathfrak N}$ must increase too.
In other words, an increase in the variables $\{\theta _i\}$ should give
rise to an increase of transport agent flow. The less is the
coefficient $R_i$, the greater is the transport agent flow that can
go through the branch $i$, all other factors being equal.  Thus, the
function $R_n(\theta )$ must be decreasing with respect to the
variable $\theta $. In addition, there should be a certain critical
concentration $\theta _c$ of the life activity products showing the
upper boundary of the vital interval for the field $\theta ({\mathbf
r},t)$, i.e. the maximum of the allowable concentration of the life
activity products in the living medium. The system has to prevent the
field $\theta ({\mathbf r},t)$ exceeding the value $\theta _c$ as much as
it can. Otherwise, the system can lose the capability for adapting.
In particular, if the field $\theta ({\mathbf r },t)$ reaches the
boundary of the vital interval at all the points of the living medium
and, consequently, all the variables $\{\theta _i\}$ come close to
the critical value $\theta _c$ (as it follows from
(\ref{*2.4})--(\ref{*2.5})) all the branches should exhaust their
capability for decreasing the coefficients $\{R_i\}$.

\begin{figure}[t]
\begin{center}
\epsfig{file=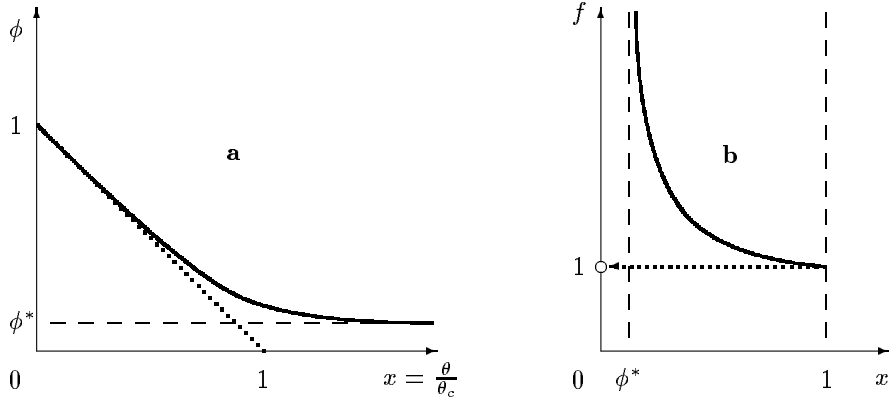}
\end{center}
\caption{Functions describing the branch response.
$($The dotted lines correspond to the ideal case.$)$}
\label{F3}
\end{figure}

Taking the aforesaid into account we represent the $R_n(\theta )$
dependence in terms of

\begin{equation}
\label{*2.7}R_n(\theta )=R_n^0\phi (\frac \theta {\theta _c}),
\end{equation}
where $R_n^0$ is a constant equal to $R_n(\theta )$ at $\theta =0$
and $ \phi (x)$ is a certain universal function of $x=\theta
/\theta _c$. The characteristic form of the $\phi (x)$~dependence is
shown in Fig.~\ref{F3}a.

For the transport network ${\mathfrak N}$ to be able to react properly to
local variations in the life activities  the transport agent flow
should be governed by branches of all the levels.  As will be seen
below this condition allows us to represent the dependence of the
value $R_n^0$ on the level number $n$ in the form

\begin{equation}
\label{*2.8}R_n^0=R_0g^n\rho (n),
\end{equation}
where $R_0$ is a certain constant and $\rho (n)$ is a smooth function
of $n$ such that $\rho (0)=1$ and formally $\rho (n)\to 0$ as $n\to
\infty $.

Concluding the description of the system response we also take into
account a possible time delay of the branch response to variations in
the variables $ \{\theta _i\}$ and represent the evolution equation
for the kinetic coefficient $R_i$ of branch $i$ belonging to level
$n$ as

\begin{equation}
\label{*2.9}\tau _n\frac{dR_i}{dt}+(R_i-R_n^0)f\Bigl( \frac{R_i}{R_n^0}
\Bigr) =-\frac{\theta _i}{\theta _c}R_n^0,
\end{equation}
where $\tau _n$ is the time delay depending solely on the level
number $n$.  This form of equation~(\ref{*2.9}) enables us to regard
the term $\theta _i/\theta _c$ as a dimensionless signal generated by
receptors imbedded into the branch $i$. As follows from (\ref{*2.7})
and (\ref{*2.9}) the functions $ f(x)$ and $\phi (x)$ are related by
the expression

\begin{equation}
\label{*3.10}[1-\phi (x)]f[\phi (x)]=x.
\end{equation}
The behavior of the function $f(x)$ is displayed in Fig.~\ref{F3}b.

In the context of the stated model the problem of self-regulation is
reduced to finding an expression relating the perfusion rate $\eta
({\mathbf r},t)$ of the transport agent to the field $\theta ({\mathbf
r},t)$:

\begin{equation}
\label{phi}\eta =\hat \Phi \{\theta \}\quad \mathrm{or}\quad \eta =\hat \Phi
\{\theta ,\eta \}.
\end{equation}
In general, this expression is of the functional form, i.e. is
nonlocal in space and only under specific conditions can relate the
fields $\eta ({\mathbf r} ,t)$, $\theta ({\mathbf r},t)$ taken at the same
point ${\mathbf r}$. In what follows we will find this conditions and
obtain the corresponding particular form of expression~(\ref{phi})

\section{\protect\label{sec:3}Governing equations for the perfusion rate.
General description}

By virtue of the adopted assumptions the size $l_N$ of the elementary
domains plays the role of an infinitesimal length. Therefore
expression~(\ref{*2.2}) can be rewritten as

\begin{equation}
\label{e3.1}\eta ({\mathbf r})=\frac 1{l_N^d}\sum_{i\in {\mathbb
L}_N}J_i\Theta _i({\mathbf r}),
\end{equation}
where the symbol $\Theta _i({\mathbf r})$ stands for the characteristic
function of the fundamental domain ${\mathfrak M}_i$, i.e.

\begin{equation}
\label{e3.2}\Theta _i({\mathbf r})=\left\{
\begin{array}{lll}
1 & \mathrm{if} & {\mathbf r}\in {\mathfrak M}_i \\
0 & \mathrm{if} & {\mathbf r}\notin {\mathfrak M}_i
\end{array}
\right.
\end{equation}
and the sum runs over all the branches of the last level. If we find
expressions relating the pattern $\{J_i\}$ to the fields $\theta
({\mathbf r},t)$, $\eta ({\mathbf r},t)$ then formula~(\ref{e3.1}) will
immediately give us the specific form of~(\ref{phi}). So in the present
section we solve the equations stated above with respect to the
variables $\{J_i\}$ regarding the field $ \theta ({\mathbf r},t)$ (and
may be the field $\eta ({\mathbf r},t)$) as fixed.

\subsection{Kirchhoff's equations}

Let us, first, find the extremal equations for functional
(\ref{*2.6}).  Following the Lagrange multiplier method we reduce the
extremum problem for functional (\ref{*2.6}) subject to conditions
(\ref{*2.4}) to finding the extremal equations of the following
functional

\begin{equation}
\label{*3.1}
{\mathbb D}_L\{J_i\}=\left[ \frac 12\sum_iR_iJ_i^2-{\mathcal E}_
{\mathrm{ext}}J_0\right] +\sum_{{\mathfrak b}}P_{{\mathfrak b}}\left[
\sum_{i\in {\mathfrak b}}^{ \mathrm{in}}J_i-\sum_{i\in {\mathfrak
b}}^{\mathrm{out}}J_i\right],
\end{equation}
where $\{P_{{\mathfrak b}}\}$ are the Lagrange multipliers ascribed
to the nodes $ \{{\mathfrak b}\}$ of the network ${\mathfrak N}$, the
sum $\sum_{{\mathfrak b}}$ runs over all the nodes $\{{\mathfrak
b}\}$, and the symbols  $\sum^{\mathrm{in}}_{i\in {\mathfrak b}}J_i$,
$\sum^{\mathrm{out}}_{i\in {\mathfrak b}}J_i$ stand
for the sum over all the branches going in or out of  the given node
${\mathfrak b}$. (The direction of the motion on the network
${\mathfrak N}$ is chosen so to coincide with the  direction of
transport agent flow.) From the conditions $\partial {\mathbb
D}_L/\partial J_i=0$ we obtain the desired extremals

\begin{equation}
\label{*3.2}J_iR_i=P_{\mathrm{in}}^i-P_{\mathrm{out}}^i
\end{equation}
and
\begin{equation}
\label{*3.3}P_{\mathrm{in}}^0-P_{\mathrm{out}}^0=\mathcal E_{\mathrm{ext}}.
\end{equation}
Here the multipliers $P_{\mathrm{in}}^i$, $P_{\mathrm{out}}^i$
correspond to the entrance and the exit of the given branch $i$  and
$P_{\mathrm{in}}^0$, $P_{ \mathrm{out}}^0$ are the multipliers
ascribed to the entrance and the exit of the transport network. It
should be noted that equations~(\ref{*3.2}) and (\ref{*3.3}) together
with equations~(\ref{*2.4}) actually make up the system of
Kirchhoff's equations for the network ${\mathfrak N}$. In these equations
the variables $\{P_{{\mathfrak b}}\}$ play the role of potentials at
the nodes $\{{\mathfrak b}\}$ causing the transport agent flow through the
corresponding branches and the kinetic coefficients $\{R_i\}$ may be
regarded as the branch resistances.

\subsection{Additional potential sources}

At the next step we should solve the system of Kirchhoff's
equations~(\ref{*2.4}), (\ref{*3.2}), (\ref{*3.3}). However, when
the field $\theta ({\mathbf r},t)$ is nonuniform, the resistances
$\{R_i\}$ of all the branches can differ in magnitude because of the
network response. In this case solving Kirchhoff's equations directly
is troublesome. In order to avoid this problem we make use of the
following trick.

Let us ascribe to each branch $i$ the quantity $\varepsilon _i$ defined by
the expression

\begin{equation}
\label{*10.8}\varepsilon _i=J_i(R_{n_i}^0-R_i),
\end{equation}
where $n_i$ is the level number of the given branch. The quantities $
\{\varepsilon _i\}$ will be called additional potential sources.
Formula~(\ref{*10.8}) allows us to rewrite equation~(\ref{*3.2}) in
terms of

\begin{equation}
\label{*10.9}J_iR_{n_i}^0=P_{\mathrm{in}}^i-P_{\mathrm{out}}^i+
\varepsilon _i.
\end{equation}
The collection of equations (\ref{*2.4}) and (\ref{*10.9}) written
for every node and every branch forms the system of Kirchhoff's
equations describing the flow pattern $\{J_i\}$ on a certain network
of the same geometry where, however, branches do not respond to
variations in $\{\theta _i\}$. In this case the self-regulation
process is effectively implemented through the appearance of the
additional potential sources $\{\varepsilon _i\}$. We will refer to
this network as the homogeneous one. In this way, when the quantities
$\{\varepsilon _i\}$ have known values the analysis of the transport
agent flow redistribution over the supplying and draining beds with
branches sensitive to $\theta _i$ is reduced to solving the
Kirchhoff's equations for the corresponding homogeneous network. In
the next subsection we will get expressions relating the additional
potential sources $ \{\varepsilon _i\}$ to the fields $\theta ({\mathbf
r},t)$, $\eta ({\mathbf r},t)$.  Therefore this trick, in fact, can be
useful in solving the original Kirchhoff's equations. In the present
subsection we obtain the expressions specifying the flow pattern
$\{J_i\}$ for fixed values of the quantities $ \{\varepsilon _i\}$.

Since the flow patterns on the supplying and draining beds are mirror
image of each other we may confine our consideration to the draining
bed only. In this case we may set $P_{\mathrm{out}}^0=0$ at the
network exit and $P_{\mathrm{in}}^i=\mathcal E_{\mathrm{ext}}/2$ for all
the branches of the last level, $i\in {\mathbb L}_N$. For the
corresponding draining bed of the homogeneous network the solution of
Kirchhoff's equations (\ref{*2.4}), (\ref{*10.9}) can be written in
the form

\begin{equation}
\label{*10.11}J_i=\sum_j\Lambda _{ij}\varepsilon _j+\frac 12\Lambda _{ii_0}
{\mathcal E}_{\mathrm{ext}}.
\end{equation}
Here $\Lambda _{ij}$ is the Green matrix, i.e. the solution of these
equations when ${\mathcal E}_{\mathrm{ext}}=0$ and $\varepsilon
_{j^{\prime }}=0$ for all branches except the branch $j$ for which
$\varepsilon _j=1$, and $ i_0$ stands for the draining bed stem. We
note that the possibility of representation (\ref{*10.11}) results
from the linearity of equations (\ref {*2.4}) and (\ref{*10.9}) with
respect to the transport agent flows $\{J_i\}$. Appendix~\ref{app:1}
specifies the Green matrix $\Lambda _{ij}$ at leading order in the
small parameter $\left[ \rho (n)-\rho (n+1)\right] /\rho (n)$ (for the
definition of $\rho (n)$ see expression~(\ref{*2.8})).

Formula (\ref{*10.11}) enables us to represent
relationship~(\ref{e3.1}) in terms of

\begin{equation}
\label{*50.9}\eta ({\mathbf r})=\frac 1{l_N^d}\sum_{i\in {\mathbb L}
_N}\Bigl(\sum_j\Lambda _{ij}\varepsilon _j+\frac 12\Lambda _{ii_0}
{\mathcal E}_{\mathrm{ext}}\Bigr)\Theta _i({\mathbf r}).
\end{equation}
This expression will lead us to (\ref{phi}) if we can describe the
dynamics of the additional potential sources $\{\varepsilon _i\}$.

\subsection{Governing equations for the additional potential sources}

Differentiating relation (\ref{*10.8}) with respect to time $t$ and
taking into account equation (\ref{*2.9}) we find the following
evolution equation for the quantity $\varepsilon _i$:

\begin{equation}
\label{*10.10}\tau _{n_i}\frac{d\varepsilon _i}{dt}+\varepsilon _i\Bigl[
f\Bigl( 1-\frac{\varepsilon _i}{J_iR_{n_i}^{\circ }}\Bigr) -\tau _{n_i}
\frac d{dt}\ln {J_i}\Bigr] =R_{n_i}^0J_i\frac{\theta _i}{\theta _c}.
\end{equation}
In obtaining (\ref{*10.10}) we also have assumed that the flow $J_i$
cannot change its direction. Due to the conservation
laws~(\ref{*2.4}) and (\ref {*2.5}) the values $J_i$ and $\theta
_iJ_i$ corresponding to the branch $i$ can be directly related to the
same variables ascribed to the branches of the subtree ${\mathfrak
N}^i$ whose stem is the given branch $i$. Passing to the last level
${\mathbb L}_N^i$ of this subtree we immediately get

\begin{eqnarray}
\label{*50.1}
J_i & = & \sum_{j\in {\mathbb L}_N^i}J_j=\int\limits_{{\mathfrak
M}}d{\mathbf r} \Theta _i({\mathbf r})\eta ({\mathbf r}),\\
\label{*50.2}J_i\theta _i & = & \sum_{j\in {\mathbb L}_N^i}J_j\theta
_j=\int\limits_{ {\mathfrak M}}d{\mathbf r}\Theta _i({\mathbf r})
\eta ({\mathbf r})\theta ({\mathbf r}),
\end{eqnarray}
where we have made us of expressions (\ref{*2.2}), (\ref{*2.3}) and
the identity

\begin{equation}
\label{*50.7}\sum_{j\in {\mathbb L}_N^i}\Theta _j({\mathbf r})=\Theta
_i({\mathbf r})
\end{equation}
which is due to~(\ref{em2}).

The system of equation~(\ref{*10.10}) and expressions~(\ref{*50.1}),
(\ref{*50.2}) describes the dynamics of the additional potential
sources in such a form that explicitly relates time variations of
$\{\varepsilon _i\}$ to the fields $\theta ({\mathbf r},t)$, $\eta ({\mathbf
r},t)$ and, thus, among with expression~(\ref{*50.9}) gives us, in
principle, the desired relationship~(\ref{phi}).

In particular, when the branches respond without delay,
i.e.  $\{\tau _n=0\} $, this system and expression~(\ref{*3.10}) gives us
the explicit relation

\begin{equation}
\label{*33}\varepsilon _i(t)=R_{n_i}^0\Bigl[
\,\int\limits_{{\mathfrak M}}d{\mathbf r} \eta ({\mathbf r},t)\Theta _i({\mathbf
r})\Bigr] \Bigl[ 1-\phi \Biggl( \frac{ \int\limits_{{\mathfrak
M}}d{\mathbf r}\eta ({\mathbf r},t)\theta ({\mathbf r},t) \Theta _i({\mathbf
r})}{\int\limits_{{\mathfrak M}}d{\mathbf r}\eta ({\mathbf r},t) \Theta
_i({\mathbf r} )}\Biggr) \Bigr].
\end{equation}
The substitution of (\ref{*33}) into (\ref{*50.9}) directly specifies the
functional dependence~(\ref{phi}).

\section{Ideal supplying network}

In this section we show that there are special conditions under which
the supplying network can function perfectly, namely, local
variations in the concentration of life activity products $\theta
({\mathbf r},t)$ cause variations in the perfusion rate $\eta
({\mathbf r},t)$ of transport agent at the same point only. Besides,
it turns out that under these conditions the response of the
supplying network keeps the concentration of life activity products
rigorously inside the vital interval. In other words, the field $
\theta ({\mathbf r},t)$ cannot go beyond the interval $(0,\theta _c)$
for a long time.

We consider a perturbation $\delta \theta ({\mathbf r},t)$ of the
field $\theta ( {\mathbf r},t)$ whose characteristic spatial scale
${\mathcal L}_\theta $ meets the inequality

\begin{equation}
\label{*4.0}{\mathcal L}_\theta \gg l_N,
\end{equation}
so we may ignore nonuniformities of the field $\theta ({\mathbf
r},t)$ as well as the field $\eta ({\mathbf r},t)$ on spatial scales
about the elementary domain size $l_N$.

\subsection{Linear model for branch response. Perfect self-regulation}

Let us find the particular form of relationship (\ref{phi}) for the
function $\phi (\theta )$ of the form

\begin{equation}
\label{*4.1}\phi ^{\mathrm{id}}(\theta )=\left\{
\begin{array}{ccc}
1-{\displaystyle\frac \theta {\theta _c}}, & \quad \mathrm{if}
& \quad 0<\theta <\theta _c\\
0, & \quad \mathrm{if} & \quad \theta >\theta _c
\end{array}
\right. .
\end{equation}
This dependence $\phi ^{\mathrm{id}}(\theta )$ as well as the
corresponding behavior of the function $f^{\mathrm{id}}(x)$ is
displayed in Fig.~\ref{F3} by dotted lines. In particular, as follows
from (\ref{*3.10}) $f^{\mathrm{id} }(x)=1$ for $0<x\leq 1$ and
$f^{\mathrm{id}}(x)$ is undetermined at $x=0$.

Besides, we adopt the following additional assumptions. First, the
delay time $\tau _n$ of the branch response is set equal for all the
branches, i.e. $\tau _n=\tau $. Second, we will ignore the term $\tau
\frac d{dt}\ln J_i$ in equation~(\ref{*10.10}). This term actually
describes dynamic interaction between additional potential sources
and has no substantial effect at least in the following two limit
cases. When $\tau \ll \tau _\theta $, where $\tau _\theta $ is the
characteristic time of variations in the living medium state, the
network response will be quasistationary, and the term $\tau \frac
d{dt}\ln J_i$ along with the first one in (\ref {*10.10}) can be
ignored. When $\tau \gg \tau _\theta $ and, in addition, the
generation rate $q$ is not too high the value of $\varepsilon _i\tau
\frac d{dt}\ln J_i$ should be negligibly small in comparison with
$\tau \frac d{dt}\ln \varepsilon _i$. Indeed, in this case
variations of the field $\theta ({\mathbf r},t)$ are too fast for the
supplying network response to follow them and the values of
$\{\varepsilon _j\}$ should be small. Since time variations of the
transport agent flow $J_i$ are due to the action of the additional
potential sources $\{\varepsilon _j\}$ we may write $J_i\propto
\{\varepsilon _j\}$ and, thus, the term $\varepsilon _i\tau \frac
d{dt}\ln J_i$ is of second order in $\{\varepsilon _j\}$. Besides,
the inequality $ \varepsilon _i<J_iR_{n_i}^0$ is considered to be
fulfilled in advance (below in the present section we will justify
this assumption) and so
$f^{\mathrm{id} }(1-\varepsilon_i/(J_iR_{n_i}^0))=1$.

Under these conditions we can rewrite equation~(\ref{*10.10}) as

\begin{equation} \label{*11.1}\tau \frac{d\varepsilon
_i}{dt}+\varepsilon _i=R_{n_i}^0\frac 1{ \theta
_c^{}}\int\limits_{{\mathfrak M}}d{\mathbf r}\Theta _i({\mathbf
r})\eta ({\mathbf r} )\theta ({\mathbf r}),
\end{equation}
where we  have also taken into account relation~(\ref{*50.2}). Then
acting by the operator $(1+$ $\tau \frac d{dt})$ on
equation~(\ref{*50.9}) and using (\ref{*11.1}) we reduce it to the
following

\begin{equation}
\label{*11.3}\tau \frac{\partial \eta }{\partial t}+\eta =\eta
_0+\frac 1{ \theta _c^{}}\int\limits_{{\mathfrak M}}d{\mathbf
r}^{\prime }\sum_j\frac 1{l_N^d} \sum_{i\in {\mathfrak L}_N}\Theta
_i({\mathbf r})\Lambda _{ij}R_{n_j}^0\Theta _j({\mathbf r}^{\prime
})\eta ({\mathbf r}^{\prime })\theta ({\mathbf r}^{\prime }).
\end{equation}
Here $\eta _0$ is the perfusion rate of transport agent through the
living medium containing no life activity products, $\theta ({\mathbf
r},t)=0$. It should be pointed out that the value $\eta _0$ may be
regarded as a certain phenomenological parameter in context of the
self-regulation problem.

By virtue of inequality~(\ref{*4.0}) we can simplify
equation~(\ref{*11.3}) in the following way. In the case under
consideration there is a spatial scale $\ell ^{*}$ such that $l_N\ll
\ell ^{*}\ll {\mathcal L}_\theta $ on which the fields $\eta
({\mathbf r},t)$, $\theta ({\mathbf r},t)$ are practically constant.
Let us choose a branch level $n^{*}\gg 1$ for which $
l_{n^{*}}\approx \ell ^{*}$ and $N-n^{*}\gg 1$. At the given level it
is possible not to distinguish between the true perfusion rate $\eta
({\mathbf r},t) $ and the perfusion rate $\eta ^{*}({\mathbf
r},t)\simeq \eta ({\mathbf r},t)$ averaged over the size $l_{n^{*}}$
of the fundamental domains ${\mathbb M}_{n^{*}}$. So we may specify
this averaging, for example, in a manner as it is done in
Appendix~\ref{app:2}, namely, $\eta ^{*}({\mathbf r},t)=
\widehat{\mathcal P} _{n^{*}}\{\eta ({\mathbf r},t)\}$. Then acting
by the operator $\widehat{P} _{n^{*}}$ on equation~(\ref{*11.3}) and
taking into account identity~(\ref {a2.av3}) we convert it to the
same form within the replacement

$$
\frac 1{l_N^d}\sum_{i\in {\mathbb L}_N}\Theta _i({\mathbf r})\Lambda
_{ij}\rightarrow \frac 1{l_{n^{*}}^d}\sum_{i\in {\mathbb
L}_{n^{*}}}\Theta _i( {\mathbf r})\Lambda _{ij}.
$$
The latter, however, enables us to make use of identity~(\ref{id})
and to reduce (\ref{*11.3}) to

\begin{equation}
\label{*11.3b}\tau \frac{\partial \eta }{\partial t}+\eta =\eta
_0+\frac 1{ \theta _c^{}}\int\limits_{{\mathfrak M}}d{\mathbf
r}^{\prime }\frac 1{l_{n^{*}}^d} \sum_{i\in {\mathbb
L}_{n^{*}}}\Theta _i({\mathbf r})\Theta _i({\mathbf r}^{\prime
})\eta ({\mathbf r}^{\prime })\theta ({\mathbf r}^{\prime }).
\end{equation}
Whence it immediately follows that

\begin{equation}
\label{psr}\tau \frac{\partial \eta }{\partial t}+\Bigl( 1-\frac \theta {
\theta _c}\Bigr) \eta =\eta _0
\end{equation}
because the fields $\theta ({\mathbf r},t)$, $\eta ({\mathbf r},t)$
are practically constant on scales of order $l_{n^{*}}$.

Equation (\ref{psr}) is the desired relationship between the
perfusion rate $ \eta ({\mathbf r},t)$ of transport agent and the
concentration $\theta ({\mathbf r} ,t)$ of life activity products.

\subsection{Characteristics of perfect self-regulation}

The local form of equation~(\ref{psr}) demonstrates that under the
given conditions the perfusion rate $\eta ({\mathbf r},t)$  is
determined by variations in the field $\theta ({\mathbf r},t)$ at the
same point only. Thus, the cooperative response of all the branches
is so self-consistent that the supplying network delivers
``nutrients'' only to the living medium points that ''ask'' for this.
In this case no unexpected changes in the ``nutrient'' delivery
interfering with the living medium activities occur, i.e. the
supplying network functions perfectly.

\begin{figure}[t]
\begin{center}
\epsfig{file=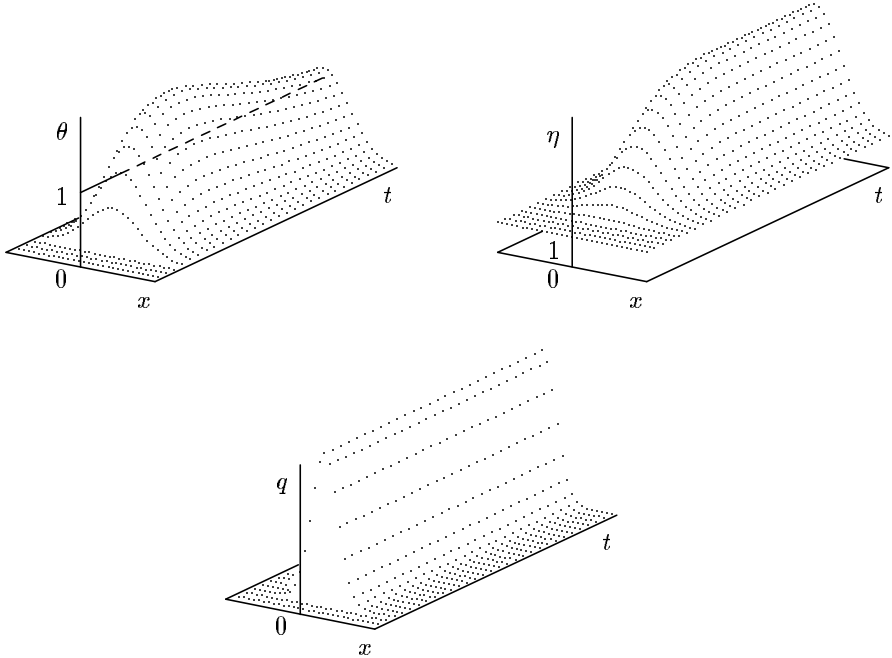}
\end{center}
%\begin{center} \input q_heat.pic \end{center}
\caption{Typical dynamics of the fields $\theta ({\mathbf r},t)$,
$\eta ({\mathbf r},t)$ under local stepwise variations in the life
activities of the living medium
$($$q({\mathbf r},t)$ is the generation rate of life activity products$)$.}
\label{F4}
\end{figure}

Another characteristics of perfect self-regulation is the fact that
the field $\theta ({\mathbf r},t)$ cannot go beyond the vital
interval $[0,\theta _c] $ for a long time. This is due to an
unbounded increase in the perfusion rate $\eta ({\mathbf r},t)$ that
occurs when the field $\theta ({\mathbf r},t)$ locally exceeds the
value $\theta _c$ (see equation~(\ref{psr})). Typical dynamics of the
fields $\theta ({\mathbf r},t)$, $\eta ({\mathbf r},t)$ demonstrating
such behavior is shown in Fig.~\ref{F4}  for the one-dimensional living
medium.

Concluding this subsection we remind that when deriving
equation~(\ref{psr}) we have used the inequality $\varepsilon
_i<J_iR_{n_i}^0$ without justification. Its validity can be shown in
the following way. Let us consider the potential drop $\Delta
P_i\equiv P_{\mathrm{in}}^i-P_{\mathrm{out} }^i$ across a branch $i$
which according to (\ref{*10.9}) is equal to $ \Delta
P_i=J_iR_{n_i}^0-\varepsilon _i$. Then taking into account
(\ref{*50.1}), (\ref{*11.1}), and (\ref{psr}) we get

\begin{equation}
\label{jus}\tau \frac{d\Delta P_i}{dt}+\Delta P_i=R_{n_i}^0\eta
_0\int\limits_{{\mathfrak M}}d{\mathbf r}\Theta _i({\mathbf r}).
\end{equation}
The right-hand side of this equation is a constant independent of
variations in the living medium state. At the initial time the field
$\theta ({\mathbf r} ,t)=0$ and so $\Delta P_i>0$, thus $\Delta P_i$
should be positive at any time moment later.

Besides, equation~(\ref{jus}) demonstrates one more characteristic
property of the perfect self-regulation. Under the given conditions
the potential distribution over the supplying network does not depend
on variations in the living medium state. Exactly this independence
ensures that the transport agent flow $J_i$ through a last level
branch $i\in {\mathbb L}_N$ will remain unchanged if the state of the
living medium inside the corresponding elementary domain ${\mathfrak
M}_i\in {\mathbb M}_N$ does not alter. In fact, for this branch the
value $\theta _i$ $\cong \theta ({\mathbf r},t)$ (${\mathbf r}\in
{\mathfrak M}_i$), so, the ``resistivity''  $R_i(\theta _i)$ of branch
$i$ will be also constant under such conditions. Consequently, the
transport agent flow $J_i=\Delta P_i/R_i(\theta _i)$ as well as the
perfusion rate $\eta ( {\mathbf r},t)=J_i/l_N^d$ in this domain will
be independent of the living medium state at other points. In other
words, it is the constancy of the potential distribution that ensures
the perfusion rate $\eta ({\mathbf r},t)$ being controlled by the
living medium state at the same point.

\section{Model generalizations}

\subsection{Dichotomic supplying network}

In the previous sections we have considered the supplying network
whose architectonics and embedding into the living medium are
specified by $g$-fold spliting of branches with $g=2^d$, where $d$
is the living medium dimension. However, looking attentively through
the previous analysis we see that it has been used, in fact, only the
following properties of the supplying network architectonics.
Each branch $i$
supplies (or drains) as a whole a certain domain ${\mathfrak  M}^i$
and all the branches of smaller length and connected with the branch
$i$ supply (or drain) subdomains of the domain ${\mathfrak  M}^i$.
The whole collection $\{{\mathfrak  M}^i\}_{i\in {\mathbb  L}_n}$ of
the domains corresponding to  branches $\{i\in {\mathbb  L}_n\}$ of
the same level ${\mathbb  L}_n$ compose together the living medim
${\mathfrak  M}$ and are mutually disjoint (see (\ref{em1})).
Besides, if $\{i\in {\mathbb  L}_n^j\}$ is the whole collection of
branches of level $n$ that are formed by splitting of the branch $j$
(of level $m\leq n$) then all the fundamental domains $\{{\mathfrak
M} ^i\}$ together make up the domain ${\mathfrak  M}^j$ (see
(\ref{em2})). None of the other properties has been used.

\begin{figure}[t]
\begin{center}
\epsfig{file=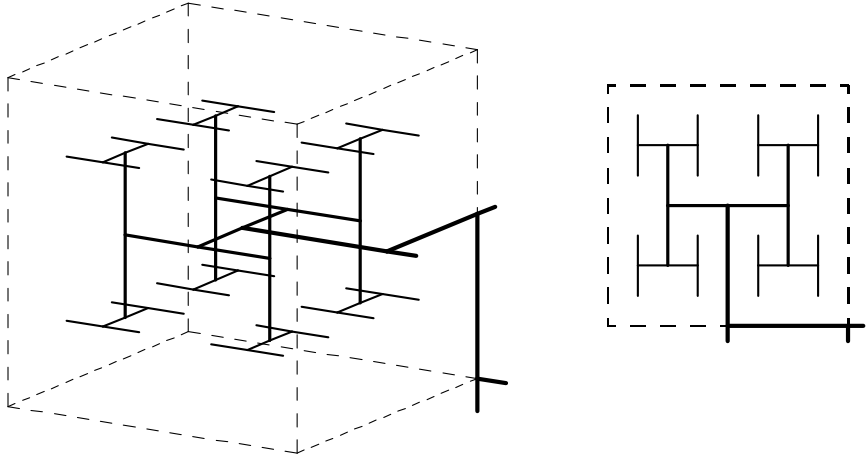}
\end{center}
\caption{Fragments of the supplying network architectonics for the
two- and three-dimensional living medium.}
\label{F5}
\end{figure}

 In particular, the factor
$g$ is not resticted by the value of $2^d$. Therefore the obtained
results hold for any other implementation of the architectonics of
the supplying network and its embedding into the living medium that
meets conditioins~(\ref{em1}), (\ref{em2}). For example, dichotomic
supplying networks demonstrated in Fig.~\ref{F5} obey these
conditions.

\subsection{Irregular supplying network}

In the previous sections we have analyzed the self-regulation problem
assuming the supplying network regular in architectonics and
involving a large number of levelvs. So it could give the impression
that these assumptions are the necessary conditions for
self-regulation to be perfect.  Therefore in the present section we
demonstrate that the perfect self-regulation can occur in systems
with supplying networks of arbitrary architectonics, at least, when
the transient processes are ignorable.

\begin{figure}[ht]
\begin{center}
\epsfig{file=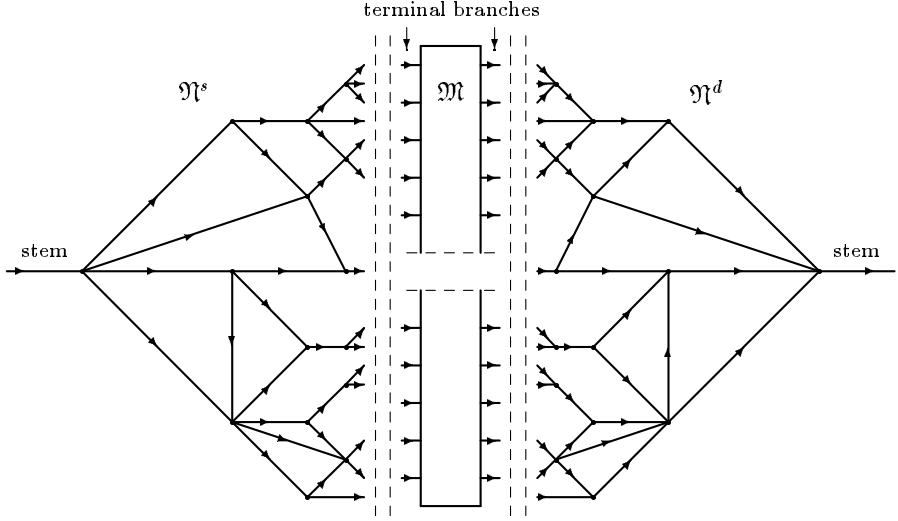}
\end{center}
\caption{Living system with irregular supplying network.}
\label{F6}
\end{figure}

As before we consider a supplying network consisting of supplying and
draining beds which are mirror image of each other and connected
through terminal branches (Fig.~\ref{F6}). The transport agent does
not interact with the living medium until it reaches the terminal
branches where the transport agent and the living medium exchange
``nutrients'' and life activity products. Then during motion through
the draining bed the transport agent does not interact with the
living medium again. Transport agent flow is assumed to be governed
by the same regularities as stated in \S\ref{sec:model}.

\begin{figure}
\begin{center}
\epsfig{file=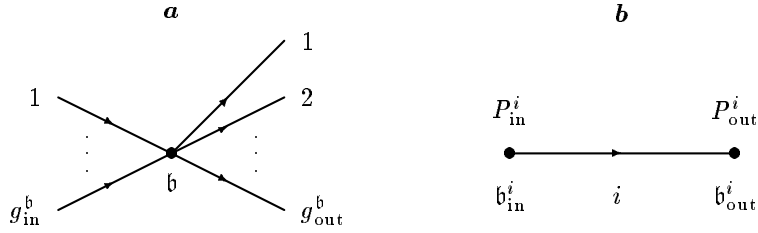}
\end{center}
\caption{Elements of the irregular supplying network.}
\label{F7}
\end{figure}

So for each node ${\mathfrak  b}$ of the draining bed
(Fig.~\ref{F7}a) we can write the following conservation laws

\begin{equation}
\label{irr1}
\sum\limits_{i\in {\mathfrak  b}}^{\mathrm
{in}}J_i=\sum\limits_{i\in {\mathfrak b}}^{\mathrm {out}}J_i\qquad
\mathrm {and}\qquad \sum\limits_{i\in {\mathfrak  b}}^{\mathrm
{in}}\theta _iJ_i=(\theta _{\mathrm {out}})_{{\mathfrak
b}}\sum\limits_{i\in {\mathfrak  b}}^{ \mathrm {out}}J_i,
\end{equation}
where the sums $\sum_{i\in {\mathfrak  b}}^{\mathrm {in}}$ and
$\sum_{i\in {\mathfrak  b}}^{ \mathrm {out}}$ run over all the
branches $i\in {\mathfrak  b}$ leading to or from this node, $J_i$
and $\theta _i$ are the transport agent flow and the concentration of
life activity products in the branche $i$ going in or out of the node
${\mathfrak  b}$. Besides, for all the branches going out of one node
$ {\mathfrak  b}$ the concentration $(\theta _{\mathrm
{out}})_{{\mathfrak  b}}$ of life activity products is assumed to be
the same. The branch orientation is chosen according to the direction
of the transport agent motion. The potential distribution
$\{P_{{\mathfrak  b}}\}$ over the draining bed nodes and the patten
$\{J_i\}$ of transport agent flow are related by the expressions

\begin{equation}
\label{irr2}J_iR_i(\theta _i)=P_{\mathrm {in}}^i-P_{\mathrm
{out}}^i\stackrel{ \mathrm {def}}{=}\Delta P_i,
\end{equation}
where $P_{\mathrm {in}}^i$ and $P_{\mathrm {out}}^i$ are the
potentials ascribed to the terminal nodes ${\mathfrak  b}_{\mathrm
{in}}^i$ and ${\mathfrak  b}_{\mathrm {out}}^i$ of the branch $i$ as
shown in Fig.~\ref{F7}b. The dependence of the ``resistance''
$R_i(\theta _i)$ of the branch $i$ on the value $\theta _i$ describes
the living system response to variations in the living medium state.
In the present section we analyze only the case corresponding to the
perfect self-regulation which is characterized by the relationship
(cf.~(\ref {*2.9}))

\begin{equation}
\label{irr3}R_i(\theta _i)=\left\{
\begin{array}{ccc}
R_i^0{\displaystyle (1-\frac{\theta _i}{\theta _c})}  & \mathrm {if} &
0<\theta _i\leq 1 \\
0  & \mathrm {if} & \theta _i\geq 1
\end{array}
\right. ,
\end{equation}
where $R_i^0$ is the ``resistance'' of the branch $i$ at $\theta
_i=0$. For the terminal branches $\{i\}^{\mathrm {term}}$ of the
draining bed the quantities $\{\theta _i\}^{\mathrm {term}}$ are
directly specified by the living medim state and, so, may be treated
as predetermined when analyzing the distribution of transport agent
flow over the draining bed. Besides, due to the mirror symmetry of
the supplying and draining beds, we can regard the potential
$P_{\mathrm {out}}^0$ at the exit node of the draining bed (the
latter node of the draining bed stem) and the potential $P_{\mathrm
{in}}^{ \mathrm {term}}$ at all the entrance nodes (the former nodes
of the terminal branches) as fixed values.

In order to show that under such condition the supplying network
response is perfect it is sufficient to demonstrate that the
potential distribution over the draining bed nodes is constant, i.e.
does not depends on variations in the field $\theta ({\mathbf r},t)$.
Indeed, in this case the perfusion rate $ \eta ({\mathbf r},t)$ of
transport agent inside a domain ${\mathfrak  M}_i^{\mathrm {term}} $
drained as a whole by the terminal branch $i^{\mathrm {term}}$ is

$$
\eta ({\mathbf r},t)\simeq \frac{J_i^{\mathrm {term}}}{V_i^{\mathrm
{term}}}=\frac{ \Delta P_i^{\mathrm {term}}}{V_i^{\mathrm
{term}}R_i(\theta _i^{\mathrm {term}})}= \frac{\const
}{R_i(\theta _i^{\mathrm {term}})},
$$
where $V_i^{\mathrm {term}}$ is the volume of the domain ${\mathfrak
M}_i^{\mathrm {term} }$. Whence it follows that the perfusion rate
$\eta ({\mathbf r},t)$ is practically controlled by the field $\theta
({\mathbf r},t)$ taken at the same point ${\mathbf r\in }{\mathfrak
M}_i^{\mathrm {term}}$ because the value of $\theta _i^{ \mathrm
{term}}$ is determined directly by the value of the field $\theta
({\mathbf r},t)$ in the domain ${\mathfrak  M}_i^{\mathrm {term}}$.

Let us fix a potential distribution $\{\tilde P_{{\mathfrak  b}}\}$
that occurs at initial time. In general the system of Kirchhoff's
equations~(\ref{irr1}), (\ref{irr2}) should have a single solution.
So to prove the independece of the potential distribution
$\{P_{{\mathfrak  b}}\}$ from variations in the field $ \theta
({\mathbf r},t)$ we may to show that this system possesses a solution
provided we have fixed $\{P_{{\mathfrak  b}}=\tilde P_{{\mathfrak
b}}\}$.

Let us consider the system of equations~(\ref{irr1}), (\ref{irr2})
for an arbitrary node ${\mathfrak  b}$ (Fig.~\ref{F7}a) which is
formed by $g_{\mathrm {in}}^{\mathfrak b}$ branches going into it and
$g_{\mathrm {out}}^{\mathfrak b}$ branches leaving this node. At
first, we will treat the values $\{\theta _i\}^{\mathrm {in}}$
ascribed to the branches going into the node ${\mathfrak  b}$ as
given quantities.  Then the variables to be found by solving this
system are as follows:  $\{J_i\}^ {\mathrm {in}}$, $\{J_i\}^{\mathrm
{out}}$, and $(\theta _{\mathrm {out}})_{{\mathfrak  b}}$ (the
subsripts $in$ and $out$ mean that the corresponding quatities relate
to the branches going into or out of the node ${\mathfrak  b}$).

The total number of these quantities is equal to $g_{\mathrm
{in}}^{\mathfrak b}+g_{\mathrm {out}}^{\mathfrak b}+1$, whereas the
total number of the given equations is $g_{\mathrm {in}}^{\mathfrak
b}+g_{ \mathrm {out}}^{\mathfrak b}+2$. Therefore this system seems
to be overdetermined. In fact, however, equations~(\ref{irr1}),
(\ref{irr2}) are linearly dependent.

To justify this statement, we, first, note that under the steady-state
conditions none of the quantities $\{\theta _i\}$ can exceed the
value $ \theta _c$. Indeed, according to (\ref{irr1}), for any node
${\mathfrak  b}$ the value of $(\theta _{\mathrm {out}})_{{\mathfrak
b}}\leq \max \{\theta _i\}_{\mathrm {in}} $. So for a branch $i$ the
value of $\theta _i$ might exceed $\theta _c$ only if there were
a path ${\mathbb  P}_i$ on the draining bed leading from the branch
$i$ to one of the terminal branches such that for all braches $\{j\}$
belonging to it the values $\theta _j\geq \theta _c$. In this case,
as follows from (\ref{irr2}) and (\ref{irr3}), the total
``resistance'' of the path ${\mathbb  P}_i$ were equal to zero. So
the potential $P_{\mathrm {out}}^i$ at the node ${\mathfrak  b}_i$
into which the branch\thinspace $i$ goes would coinside with the
potential at the draining bed entrance, $P_{\mathrm {out}}^i$ $=$
$P_{ \mathrm {in}}^{\mathrm {term}}$. In addition, the total transport
agent flow $ \sum_{i\in \mathfrak b_i}^{\mathrm {out}}J_i$
going through the
node ${\mathfrak  b}_i$ would be equal to the transport agent flow
along the path ${\mathbb  P}_i$ except for cases when there are
another similar paths ${\mathbb P}_i^{\prime }$. This in turn would
cause the value of $(\theta _{\mathrm {out}})_{{\mathfrak  b}_i}$ to
become greater than (or equal to) $\theta _c$ and so on. In this way we
would reach the stem of the draining bed and obtain that the potential
$P_{\mathrm {out}}^0$ at the exit of the draining bed has to be equal
to the portential $P_{\mathrm {in}}^{ \mathrm {term}}$ at its
entrance. Thus all the values $\theta _i$ do not exceed $\theta _c$.

Then let us sum equations (\ref{irr2}) with the wight $1/R_i^0$ over
all branches relating to the node ${\mathfrak  b}$. In this way
taking into account (\ref{irr1}) and (\ref{irr3}) we get

\begin{equation}
\label{irr4}\sum\limits_{i\in {\mathfrak  b}}^{\mathrm {in}}\frac{\tilde P_{\mathrm {in}
}^i-\tilde P_{{\mathfrak  b}}}{R_i^0}=\sum\limits_{i\in {\mathfrak  b}}^{\mathrm {out}}\frac{
\tilde P_{{\mathfrak  b}}-\tilde P_{\mathrm {out}}^i}{R_i^0}.
\end{equation}
Since the initial portential distribution $\{\tilde P_{{\mathfrak
b}}\}$ has been established by the existing transport agent flow this
equality is fulfilled.  So equations~(\ref{irr1}), (\ref{irr2}) are
linearly dependent and can be reduced to a system of $(g_{\mathrm
{in}}^{\mathfrak b}+g_{\mathrm {out}}^{\mathfrak b}+1)$ equations
which is not overdetermined and possesses a solution.

Above we have regarded the quantities $\{\theta _i\}$ as given
beforehand.  If a node ${\mathfrak  b}$ is one of the latter nodes of
the terminal branches this is justified because the values $\{\theta
_i\}$ at such branches are directly determined by the living medium
state. Then, sovling the system of
equations~(\ref{irr1})--(\ref{irr2}) for these nodes we find the
values $ \{(\theta _{\mathrm {out}})_{{\mathfrak  b}}\}$ which form
the collection of quatities $\{\theta _i\}_{\mathrm {int}}$ for nodes
of lower levels. Repeating this proceedures we get the draining bed
stem.

In this way we get the conclusion that there is a solution of the
full set of equation~(\ref{irr1})--(\ref{irr2}) for a fixed potential
distribution $ \{P_{{\mathfrak b}}\}$ when the state of the living
medium varies in time. The latter demonstrates that under the adopted
assumptions the response of the supplying network is perfect.

\section{Essence of perfect self-regulation}

The fact that equation~(\ref{psr}) governing the dynamics of
transport agent perfusion can be of such a simple form is a surprise
because it does not directly contain any information on the complex
architectonics of the supplying network. This fact is actually one of
the main results obtained in the present paper. So we, now, discuss
the qualitative physical essence of haw the perfect regulation can
occur.

\begin{figure}
\begin{center}
\epsfig{file=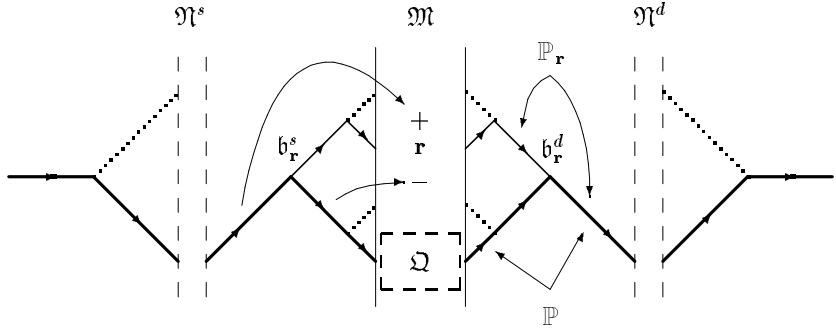}
\end{center}
\caption{Illustration of the mechanism by which the
nonlocal dependence of the perfusion rate  $\eta\{\theta\}$ on the
field $\theta (\mathbf r)$ at distant points is suppressed.}
\label{F8}
\end{figure}

Let us assume that in a small domain $\mathfrak Q$ (Fig.~\ref{F8}) the field
$\theta ( {\mathbf r},t)$ increases and comes close to the boundary
$\theta _c$ of the vital interval. In order to smother this dangerous
growth the system should increase the perfusion rate $\eta ({\mathbf
r},t)$ in the domain $\mathfrak Q$. For this purpose all the ``resistances''
$\{R_i\}_{{\mathbb P}}$ along the path ${\mathbb P} $ connecting the
stems of the supplying and draining beds and going through the domain
$\mathfrak Q$ have to decrease substantially. As a result, the transport agent
flow along the path ${\mathbb P}$ and, consequently, the perfusion
rate in the domain $\mathfrak Q$ will grow. Higher values of the variables
$\{\theta _i\}_{ {\mathbb P}}$ along the path ${\mathbb P}$ (on the
draining bed) bear the information required of such a response.

Variations in the field $\theta ({\mathbf r},t)$ located in the
domain $\mathfrak Q$ can, in principle, give rise to alterations of the
perfusion rate at points external to the domain $\mathfrak Q$ because of
variations in the transport agent flow through large branches.
However there is a difference between the external and internal
points of the domain $\mathfrak Q$ from the standpoint of
this transport agent
flow redistribution. For an external point ${\mathbf r}\notin\mathfrak Q$
on the path ${\mathbb P}$ there are two nodes ${\mathfrak
b}_{{\mathbf r}}^s$ and ${\mathfrak b}_{{\mathbf r}}^d$ belonging to the
supplying and draining beds, respectively, at which a similar path
${\mathbb P}_{{\mathbf r}}$ leading to the given point ${\mathbf r}$,
at first, diverges from the path ${\mathbb P}$ and, then, they
converge (Fig.~\ref{F8}).  Due to the mirror symmetry of the two beds
we can confine our consideration to the supplying bed only. The node
${\mathfrak b}_{{\mathbf r}}^s$ divides the path $ {\mathbb P}$ on
this bed into two parts. One of them leads from the stem to the node
$b_{{\mathbf r}}^s$ and coincides with the path ${\mathbb
P}_{{\mathbf r}}$. The other goes from the node ${\mathfrak
b}_{{\mathbf r}}^s$ to the domain $\mathfrak Q$ and does not belong to the
path ${\mathbb P}_{{\mathbf r}}$. Therefore, on one hand, decrease in the
``resistances'' of the branches forming the former part tends to
increase the transport agent flow along the path ${\mathbb P}$ and,
thus, through small branches of the path ${\mathbb P}_{{\mathbf r}}$
that directly control the perfusion rate in the vicinity of the point
${\mathbf r}$. On the other hand, the less are the ``resistances'' of
the branches making up the latter part of the path ${\mathbb P}$, the
larger is the fraction of transport agent flow that is directed along
the path ${\mathbb P}$ at the node ${\mathfrak b}_{ {\mathbf r}}^s$.
These effects are opposite in sign with respect to increase in the
perfusion rate at the point ${\mathbf r}$ as illustrated in
Fig.~\ref{F8} and can compensate each other. In the present paper we
have actually found the conditions of such compensation.

\section{Conclusion}

In the present paper we have shown that there is a specific
cooperative mechanism of self-regulation by which a living system can
respond perfectly to changes in the environment, i.e. react without
interference with itself.  This mechanism is based on the
corresponding self-processing of information and the cooperative
effect caused by individual response of the supplying network
elements. The existence of large hierarchical systems in nature and
their capability for adapting to changes in the environment points to
the fact that they should function perfectly, at least, at first
approximation.  For example, in living tissue in order to reduce
effects of nonideality arterial and venous beds contain a system of
anastomoses, i.e. vessels joining arteries or veins of the same level
\cite{2}. In other words, the vascular network is organized in such a
manner that its function be as perfect as possible.

So we think the obtained results will be useful in analysis of real
living systems and can form the basis of the following mathematical
modelling of complex self-regulation processes.

\section*{Acknowledgements}

The authors would like to thank Yu.~A.~Danilov and A.~V.~Priezzhev
for attention to the work and useful criticisms.

\appendix
\section{The Green matrix $\Lambda _{ij}$\label{app:1}}

\begin{figure}
\begin{center}
\epsfig{file=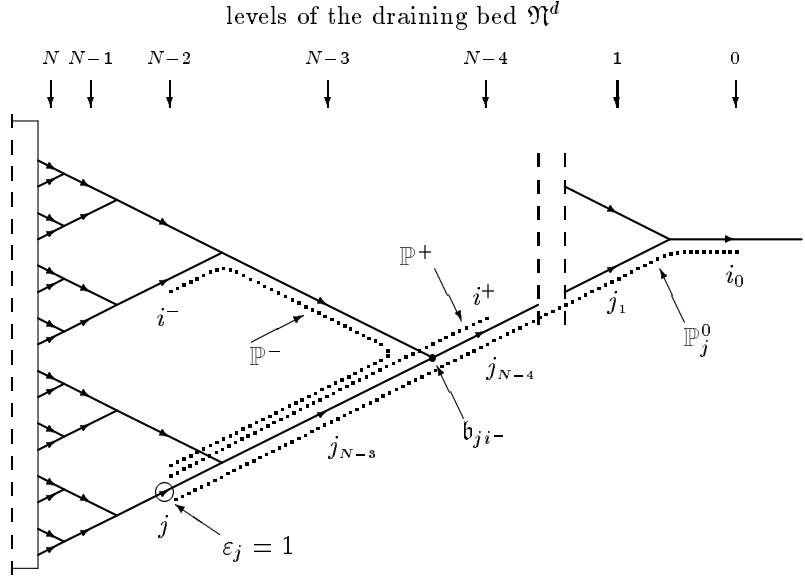}
\end{center}
\caption{Homogeneous draining bed with a single potential source of unit
power, $\varepsilon _j = 1$, located at the branch $j$. $($For simplicity
is drawn the binary draining bed, $g = 2$.$)$}
\label{Fa1}
\end{figure}

By definition, the Green matrix $\Lambda _{ij}$ gives the flow
pattern $ \{J_i=\Lambda _{ij}:j\;$is fixed$\}$ on the homogeneous
draining bed (Fig.~\ref{Fa1}) provided the branch $j$ is fixed and
contains a single potential source of power $\varepsilon _j=1$.
(For simplicity of drawing we depict the binary draining  bed in
Fig.~\ref{Fa1}.) In other words, as results from (\ref {*2.4}) and
(\ref{*10.9}) the Green matrix $\Lambda _{ij}$ is the solution of the
following system of the conservation equations at the nodes $\{{\mathfrak
b} \}$

\begin{equation}
\label{a1.1}\sum\limits_{i\in {\mathfrak  b}}^{\mathrm{in}}\Lambda
_{ij}=(\Lambda _{ij})_{\mathrm{out}}^{{\mathfrak  b}},
\end{equation}
and the potential equations

\begin{equation}
\label{a1.2}\Lambda
_{ij}R_{n_i}^0=P_{\mathrm{in}}^i-P_{\mathrm{out}}^i+\delta _{ij}.
\end{equation}
Here the sum runs over all the branches $i\in {\mathfrak  b}$ going
into the node $ {\mathfrak  b}$ and $\delta _{ij}$ is the Kronecker
symbol ($\delta _{ij}=1$ if the branches $i$, $j$ coincide with each
other and $\delta _{ij}=0$, otherwise).

In order to specify the Green matrix $\Lambda _{ij}$ we classify all
the possible pairs of branches $\{i,j\}$ into two groups. The first
group comprises all pairs $\{i,j\}_{+}$ that can be joined by a path
on the draining bed directed either from higher to lower levels or
vice versa, i.e.  by a path of constant direction. Such a path is
depicted in Fig.~\ref{Fa1} by the dotted line ${\mathbb  P}^{+}$. The
second group involves the pairs $ \{i,j\}_{-}$ that can be joined by
a path whose direction changes at a certain node ${\mathfrak
b}_{ij}$ as shown in Fig.~\ref{Fa1} by the dotted curve $ {\mathbb
P}^{-}$. In other words, this path initially goes, for example, from
the branch $j$ towards the stem until it reaches the node ${\mathfrak
b}_{ij}$ (in Fig.~\ref{Fa1} it is the node $\mathfrak b_{ji^-}$) and then
goes towards the branch $i$ in the opposite direction. To the node $
{\mathfrak b}_{ij}$ we also ascribe the level number $n_{ij}$ of branches
going into it. The given classification allows us to state the following.

\begin{proposition}
Let us introduce the quatities
$\{Z(n)\}$ $(n=0,1,\ldots ,N)$ such that:

\begin{equation}
\label{a1.5}Z(n)=\sum_{m=n}^N\rho (m),
\end{equation}
where the function $\rho (m)$ is defined by formula~(\ref{*2.8}).
Then at leading order in the small parameter $\rho (n)/Z(n)$
the Green matrix $\Lambda _{ij}$ is specified by the expression for the
$\{i,j\}_{+}$ pairs

\begin{subequations}\label{Gr1}
\begin{eqnarray}
\label{Gr1a}\Lambda _{ij} & = &\frac 1{R_0}g^{-n_j}\frac 1{Z(n_i)}\quad
\mathrm{if} \quad n_i\leq n_j,\\
\label{Gr1b}\Lambda _{ij} & = &\frac 1{R_0}g^{-n_i}\frac 1{Z(n_j)}\quad
\mathrm{if} \quad n_i\geq n_j,
\end{eqnarray}
\end{subequations}
and for the $\{i,j\}_{-}$ pairs

\begin{equation}
\label{Gr2}\Lambda _{ij}=-\frac 1{(g-1)R_0}g^{n_{ij}-n_i-n_j}\frac{\rho
(n_{ij})}{[Z(n_{ij})]^2},
\end{equation}
where $n_i$ is the level number of branch $i$.
\end{proposition}

The given proposition is the main result of the present Appendix and the
remaining part will be devoted to its substantiation.

\begin{proof}
Let us, first, prove this statement in the case when the branch $j$
is the stem $i_0$. The values of $\Lambda _{ii_0}$ are the same for
all the branches $\{i:i\in {\mathbb  L}_{n_i}\}$ belonging to one
level ${\mathbb  L}_{n_i}$.  So from (\ref{a1.1}) we get

\begin{equation}
\label{a1.3}\Lambda _{ii_0}=g^{-n_i}\Lambda _{i_0i_0}
\end{equation}
because level ${\mathbb  L}_n$ involves exactly $g^n$ branches. Then
choosing any path ${\mathbb  P}$ on the draining bed leading form the
stem $i_0$ to a branch $i_N$ of the last level and summing
equations~(\ref{a1.2}) along this path we obtain

\begin{equation}
\label{a1.4}\sum_{i\in {\mathbb  P}}\Lambda _{ij}R_{n_i}^0\equiv \Lambda
_{i_0i_0}R_0Z(0)=1,
\end{equation}
where we have taken into account expressions$~$(\ref{*2.8}) and
(\ref{a1.3}). Whence we immediately find the expression for $\Lambda
_{i_0i_0}$ and from (\ref{a1.3}) get the desired formula

\begin{equation}
\label{a1.6}\Lambda _{ii_0}=g^{-n_i}\frac 1{R_0Z(0)},
\end{equation}
which gives the same result as does formulae~(\ref{Gr1}) (namely,
(\ref{Gr1b}) for $n_{i_0} = 0$) because all the pairs $\{ii_0\}$ belong
to the first group $\{ij\}_{+}$.

\begin{figure}
\begin{center}
\epsfig{file=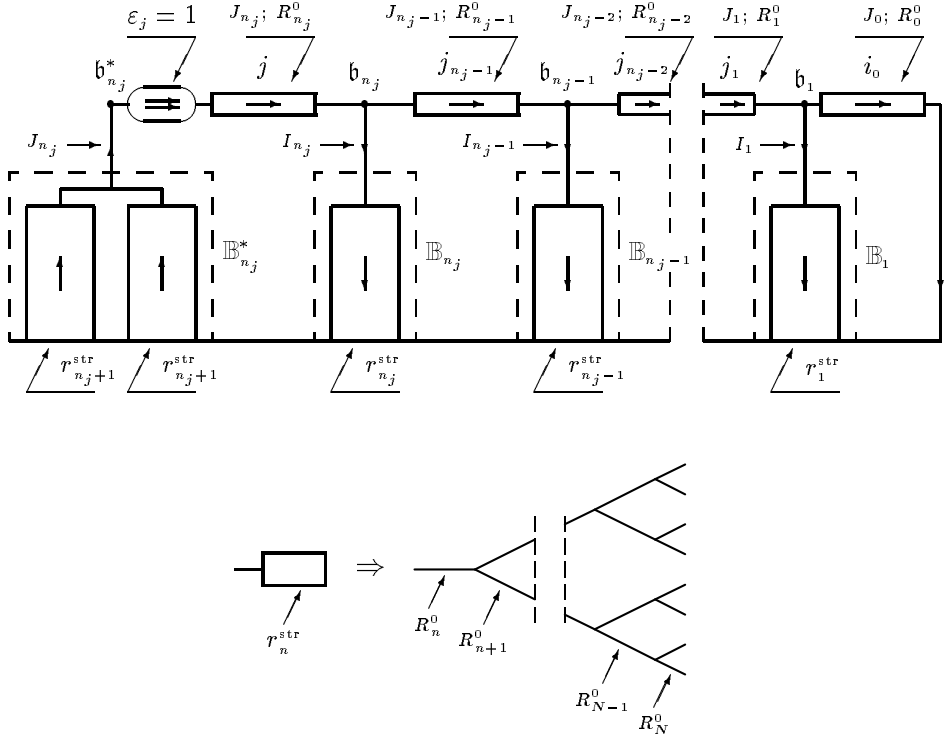}
%\input fig2app.pic
%\begin{center}
%\input fig3app.pic
\end{center}
\caption{The block representation $($upper graph$)$ of the
homogeneous draining bed equivalent to the graph shown in
Fig.~\protect\ref{Fa1} and the block structure $($lower graph$)$.
$($A block unites the draining bed subtrees characterized by identical
transport agent flow distribution and $r^{\mathrm{str}}_n$ is the total
``resistance'' of the corresponding subtree.$)$}
\label{Fa2}
\end{figure}

To find $\Lambda _{ij}$ for a branch $j$ whose level number $n_j\geq
1$ we consider a path ${\mathbb  P}_j^0$ on the network shown in
Fig.~\ref{Fa1} by the dotted line that goes from the given branch $j$
to the stem $i_0$. This
path divides the draining bed into the path ${\mathbb P}_j^0$
itself and disjoint subtrees whose stems are connected with the path
${\mathbb  P}_j^0$ through its nodes. The values of $ \Lambda
_{ij}$ for all branches of one level belonging to the same subtree
are equal. This allows us to transform the graph shown in
Fig.~\ref{Fa1} into one shown in Fig.~\ref{Fa2}, where the path
${\mathbb P}_j^0$ is represented by the sequence of branches
designated by $ \{j,j_{n_j-1},j_{n_j-2},\ldots ,j_1,i_0\}$. Here the
symbol $j_n$ stands for the branch of level $n$ belonging to this
path, with $j_{n_j}$ and $j_0$ indicating the same as $j$ and $i_0$,
respectively. The kinetic coefficients corresponding to these
branches are $\{R_{n_j}^0,\,R_{n_j-1}^0, \,R_{n_j-2}^0,\ldots
,\,R_1^0,\,R_0^0\}$. These branches at the terminal points (nodes
bound up with their entrances and exits) $\{{\mathfrak  b}_{n_j}^{*},
{\mathfrak  b}_{n_j},{\mathfrak  b}_{n_j-1},\ldots ,{\mathfrak
b}_1\}$ are connected with the blocks $\{{\mathbb
B}_{n_j}^{*},{\mathbb  B}_{n_j},{\mathbb  B}_{n_j-1},\ldots ,{\mathbb
B} _1\}$ of identical subtrees. These blocks except for
the former one $ {\mathbb  B}_{n_j}^{*}$ $($connected with the node
${\mathfrak  b}_{n_j}^{*})$ involve $ (g-1)$ subtrees, the block
${\mathbb  B}_{n_j}^{*}$ contains $g$ subtrees. The transport agent
flow spreads over such subtrees uniformly.
Therefore each block, for example, ${\mathbb  B}_n$ connected with a
node ${\mathfrak  b}_n$ can be treated as a single element
characterized by the kinetic coefficient (``resistance'')

\begin{equation}
\label{a1.7}r_n = \frac{P_{{\mathfrak  b}_n}}{I_n}.
\end{equation}
Here $I_n$ is the total flow of transport agent through the given blook,
$P_{\mathfrak b_n}$ is the potential at the node ${\mathfrak  b}_n$
induced by the potential source $\varepsilon _j=1$, and for the sake
of simplicity we have set the potential at the entrances of the last
level branches equal to zero, $P_{ \mathrm{in}}^N=0$. As is seen
from the diagram shown in Fig.~\ref{Fa2}~(lower graph), for $ 1\leq n\leq
n_j$ the value of $r_n$ is specified by the expression (see also
Fig.~\ref{Fa2})

\begin{eqnarray}
\nonumber
r_n & = & \frac 1{(g-1)} r_n^{\mathrm{str}} =
\frac 1{g-1}R_n^0+\frac 1{(g-1)g}R_{n+1}^0+\frac
1{(g-1)g^2} R_{n+2}^0+\ldots\\
\label{a1.8}
{} & = & \frac1{(g-1)}\sum_{p=n}^Ng^{-(p-n)}R_p^0
\end{eqnarray}
because level $p\geq n$ contains $(g-1)g^{p-n}$ identical branches
belonging to this block. Expressions ~(\ref{*2.8}) and (\ref{a1.5})
enable us to rewrite (\ref{a1.8}) as

\begin{equation}
\label{a1.9a}r_n=\frac 1{(g-1)}g^nR_0Z(n).\;
\end{equation}
In a similar way we obtain the expression for the kinetic coefficient
of the last block connected with the path ${\mathbb P}_j^0$
through the node ${\mathfrak  b} _{n_j}^{*}$

\begin{equation}
\label{a1.9b}r_{n_j+1}=g^{n_j}R_0Z(n_j+1).
\end{equation}

Let us introduce the quantities
$$
J_n\stackrel{\mathrm{def}}{=}\Lambda _{j_nj}\quad \mathrm{and}\quad
I_n\stackrel{ \mathrm{def}}{=}\sum_{j_{\mathrm{stem}}\in {\mathfrak
b}_n}\Lambda _{j_{\mathrm{stem}}j}
$$
being the transport agent flow through a branch $j_n$ and the total
transport agent flow in the subtree block ${\mathbb  B}_n$,
respectively, which are induced by the potential source $\varepsilon
_j=1$. In the last expression the sum runs over all the stems of the
subtrees forming the block ${\mathbb  B}_n$. Then the given reduction
of the initial graph to the subtree blocks (Fig.~\ref{Fa2})
enables us to rewrite the Kirchhoff equations~(\ref{a1.1}),
(\ref{a1.2}) in the form for $1\leq n\leq n_j$:

\begin{equation}
\label{a1.10}I_n=J_n-J_{n-1}
\end{equation}
and for $2\leq n\leq n_j$

\begin{equation}
\label{a1.11}I_nr_n=J_{n-1}R_{n-1}^0+I_{n-1}r_{n-1}.
\end{equation}
These equations are completed by the following expressions bound up
with the first and the last elements of the given network:

\begin{equation}
\label{a1.12}I_1r_1=J_0R_0\;,
\end{equation}
\begin{equation}
\label{a1.13}J_{n_j}[r_{n_j+1}+R_{n_j}^0]+I_{n_j}r_{n_j}=1\;.
\end{equation}
Due to $\rho (n)$ being a smooth function of $n$ the ratio $\rho
(n)/Z(n)$ may be treated as a small value for $N-n\gg 1$. The latter
allows us to find directly the solution of
equations~(\ref{a1.10})--(\ref{a1.13}), which is matter of the next
step.

\textbf{Continuous solution of
equations~(\protect\ref{a1.10})--(\protect\ref{a1.13})}
Taking into account (\ref{*2.8}), (\ref{a1.5}), (\ref{a1.9a}), and
(\ref {a1.9b}) we can rewrite equation~(\ref{a1.11})--(\ref{a1.13})
in terms of

\begin{equation}
\label{a1.11n}gI_nZ(n)=(g-1)J_{n-1}\rho (n-1)+I_{n-1}Z(n-1)
\end{equation}
for $2\leq n\leq n_j$ and

\begin{equation}
\label{a1.12n}gI_1Z(1)=(g-1)J_0,
\end{equation}
\begin{equation}
\label{a1.13n}J_{n_j}+\frac 1{(g-1)}I_{n_j}=\frac 1{R_0Z(n_j)}g^{-n_j},
\end{equation}
where, in addition, we have used the identity $Z(n+1)+\rho (n)=Z(n)$
and the equality $\rho (0)=1$.

Let us analyze, at first, the solution of equations (\ref{a1.10}) and
(\ref {a1.11n}) when $n\sim 1$. For such values of $n$ the function
$\rho (n)$ as well as $Z(n)$ may be considered as being constant,
$\rho (n)\cong \rho (0)=1 $ and $Z(n)\cong Z(0)\gg 1$ due to
smoothness of the function $\rho (n) $. We seek the solution of these
equations in the form

\begin{equation}
\label{a1.s}J_n=A_{+}\zeta _{+}^n+A_{-}\zeta _{-}^n,
\end{equation}
where $A_{+,-};\zeta _{+,-}$ are some constants. Substituting
(\ref{a1.s}) into (\ref{a1.10}) we get

\begin{equation}
\label{a1.si}I_n=A_{+}\zeta _{+}^{(n-1)}(\zeta _{+}-1)+A_{-}\zeta
_{-}^{(n-1)}(\zeta _{-}-1)
\end{equation}
Then the substitution of this expression into (\ref{a1.11n}) shows us
that the constants $\zeta _{+},\zeta _{-}$ are the roots of the
equation

\begin{equation}
\label{a1.r}(\zeta -1)(g\zeta -1)=\frac{g-1}{Z(0)}\zeta .
\end{equation}
Whence it follows that to the first order in the small parameter
$1/Z(0)$:

\begin{equation}
\label{a1.zeta}\zeta _{+}=1+\frac 1{Z(0)},\quad \zeta _{-}=\frac 1g\left( 1-
\frac 1{Z(0)}\right) .
\end{equation}
To the same order in $1/Z(0)$ from (\ref{a1.10}) and (\ref{a1.12n})
we find that the constants $A_{+},A_{-}$ are related by the
expression

\begin{equation}
\label{a1.ratio}\frac{A_{-}}{A_{+}}=\frac 1{(g-1)Z(0)}\ll 1.
\end{equation}

According to (\ref{a1.zeta}) the first terms on the right-hand side
of (\ref {a1.s}) and (\ref{a1.si}) are increasing functions of $n$,
whereas the second ones are decreasing functions. Therefore, by
virtue of (\ref{a1.ratio}) in the two expressions the former term is
 substantially greater than the latter for all $n\sim 1$. So, we may
ignore the second term and regard the first one as a function of the
continuous variable $n$ because of $(\zeta _{+}-1)\ll 1$. This allows
us to set

\begin{equation}
\label{a1.20}J_n-J_{n-1}=\frac{\partial J_n}{\partial n}
\end{equation}
in equation (\ref{a1.10}) and $I_{n-1}\approx I_n$ in (\ref{a1.11n}),
and also to rewrite (\ref{a1.s}) in terms of

\begin{equation}
\label{a1.21}J_n\simeq A_{+}\exp \Bigl\{ \frac n{Z(0)}\Bigr\} \;.
\end{equation}
In this case the system of equations (\ref{a1.10}) and (\ref{a1.11n})
is reduced to the equation

\begin{equation}
\label{a1.22}\frac{\partial J_n}{\partial n}=\frac 1{Z(0)}J_n
\end{equation}
and function (\ref{a1.22}) is its general solution. So, the
``boundary condition''~(\ref{a1.12n}) is responsible only for the
existence of the second terms in (\ref{a1.s}), (\ref{a1.si}) and,
therefore, can be ignored.

As follows from equations (\ref{a1.10}), (\ref{a1.11})
(or~(\ref{a1.11n})) their general solution contains two arbitrary
constants, for example, $J_0$, $J_1$. Indeed, all the other values of
$I_n$ and $J_n$ can be found by iteration. The same situation is met
with respect to the general solution~(\ref{a1.s}), (\ref{a1.si}) of
this equations for $n\sim 1$. So there is no solution of equations
(\ref{a1.10}), (\ref{a1.11}) different from (\ref{a1.s}),
 (\ref{a1.si}) for such values of $n$. Thereby, the influence of the
``boundary condition''~(\ref{a1.12}) on the solution of equations
(\ref {a1.10}) and (\ref{a1.11}) is ignorable and we may seek this
solution in the class of smooth functions of the continuous variable
$n$. In this case it contains only one arbitrary constant specified
by the ``boundary condition''~(\ref{a1.13}) (or~(\ref{a1.13n})).

Within the framework of the adopted assumptions the quantities $J_n$
and $ I_n $ as well as $\rho (n)$ and $Z(n)$ are related by the
expressions

\begin{equation}
\label{a1.23}I_n=\frac{\partial J_n}{\partial n},
\end{equation}
\begin{equation}
\label{a1.24}\rho _n=-\frac{\partial Z(n)}{\partial n},
\end{equation}
and the system of equations (\ref{a1.10}), (\ref{a1.11n}),
(\ref{a1.13n}) is reduces to the following equation

\begin{equation}
\label{a1.25}\frac{\partial J_n}{\partial n}Z(n)+J_n\frac{\partial Z(n)}
{\partial n}=0
\end{equation}
subject to the boundary condition

\begin{equation}
\label{a1.26} J_{n_j}=g^{-n_j}\frac 1{R_0Z(n_j)}.
\end{equation}
In obtaining these expressions we have also taken into account
relation~(\ref {a1.10}) and the inequalities $\left|
I_{n+1}-I_n\right| \ll I_n$ and $ \left| Z(n+1)-Z(n)\right| \ll
Z(n)$. Besides, we have ignored the second term on the left-hand side
of (\ref{a1.13n}) because of $ I_{n_j}=J_{n_j-1}-J_{n_j}\ll J_{n_j}$.
The solution of equation~(\ref{a1.25}) meeting the boundary
condition~(\ref{a1.26}) is of the form

\begin{equation}
\label{a1.27}J_n=g^{-n_j}\frac 1{R_0Z(n)}
\end{equation}
and, thus, from (\ref{a1.23}) and (\ref{a1.24}) we obtain
\begin{equation}
\label{a1.28}I_n=g^{-n_j}\frac{\rho (n)}{R_0[Z(n)]^2}.
\end{equation}

For a branch $i$ located on the path ${\mathbb P}_j^0$, i.e. $i\in
\{j,j_{n_j-1},j_{n_j-2},\ldots ,j_1,i_0\}$ the pair $\{ij\}$ belongs
to the first group and $\Lambda _{ij}=J_{n_i}$ ($n_i$ is the level
number of the branch $i$). Whence we directly get
formula~(\ref{Gr1a}).  If a branch $i$ belongs to the last block
 ${\mathbb  B}_{n_j}^{*}$ of subtrees it level number $n_i<n_j$ and
the pair $\{ij\}$ also belongs to the first group. Since for a fixed
value of $n_i$ the total number of such branches is $g^{n_i-n_j}$ and
the transport agent flow ($\Lambda _{ij}$) is the same for all of
them, we can write $\Lambda _{ij}=J_{n_j}/g^{n_i-n_j}$, which together
with expression~(\ref{a1.27}) gives equality~(\ref{Gr1b}).

Now let us consider a branch $i$ located in one of the blocks
$\{{\mathbb  B} _{n_j},{\mathbb  B}_{n_j-1},\ldots ,{\mathbb
B}_1\}$, e.g. in a block ${\mathbb  B}_{n_{ {\mathfrak  b}}}$. For
this branch the pair $\{ij\}$ belongs to the second group and the
node ${\mathfrak  b}$ at which the block ${\mathbb  B}_{n_{{\mathfrak
b}}}$ joins the path $\mathbb P_j^0$ plays the role of the node ${\mathfrak
b}_{ij}$ of the pair $ \{ij\}_{-}$. For a fixed level $n_i\geq
n_{{\mathfrak  b}}$ the total number of such branches is
$(g-1)g^{n_i-n_{{\mathfrak  b}}}$ and the total transport agent flow
going through this block is $I_{n_{{\mathfrak  b}}}$. So in this case
$\Lambda _{ij}=-I_{n_B}/[(g-1)g^{n_i-n_{{\mathfrak  b}}}]$, whence
taking into account (\ref {a1.28}) we get formula~(\ref{Gr2}). In
this expression the sign ``--'' means that in the given branch $i$
the transport agent flow $\Lambda _{ij}$ caused by the additional
potential source $\varepsilon _j=1$ is directed from lower to higher
levels. In other words, this component of the transport agent flow
through the branch $i$ and the total transport agent flow $J_i$
induced by the collective action of all the potential sources have
opposite directions.

In this way we have considered all the branches, so, proven the stated
proposition.\qquad\end{proof}

\section{The embedding identities\label{app:2}}

Let us specify the averaging procedure ${\hat \mathcal P}_n:$ $\Psi
({\mathbf  r} )\rightarrow \tilde \Psi ({\mathbf  r})$ of a function
$\Psi ({\mathbf  r})$ over spatial scales of the fundamental domains
$\{{\mathfrak  M}_n\}$ of level $n$ by the expression

\begin{equation}
\label{a2.av1}\widehat{P}_n\left\{ \Psi ({\mathbf  r)}\right\}
\stackrel{ \mathrm{def}}{\mathbf  =}\int\limits_{{\mathfrak
M}}d{\mathbf  r}^{\prime }P_n({\mathbf  r},{\mathbf  r}^{\prime
})\Psi ({\mathbf  r}^{\prime }),
\end{equation}
where the kernel

\begin{equation}
\label{a2.av2}P_n({\mathbf  r},{\mathbf  r}^{\prime })=\frac 1{l_n^d}\sum_{i\in {\mathbb  L
}_n}\Theta _i({\mathbf  r})\Theta _i({\mathbf  r}^{\prime })
\end{equation}
and the sum runs over all the branches $\{i\}$ forming the level
${\mathbb  L}_n$, for example, of the draining bed. Then we can state
the followig.

\begin{proposition}
Let ${\mathbb  L}_n$ and ${\mathbb  L}_m$ be the branch collections
forming the levels $n$ and $m$, respectively, and $n\geq m$.  Then

\begin{equation}
\label{a2.av3}\frac 1{l_m^d}\sum_{i\in {\mathbb  L}_m}\Theta
_i({\mathbf  r})\Lambda _{ij}=\widehat{P}_m\Biggl\{ \frac
1{l_n^d}\sum_{i\in {\mathbb  L}_n}\Theta _i( {\mathbf  r})\Lambda
_{ij}\Biggr\} .
\end{equation}
\end{proposition}

\begin{proof}
We prove identity (\ref{a2.av3}) by the direct substitution of
(\ref{a2.av2}) into the right-hand side of (\ref{a2.av3}). In this
way we get

\begin{eqnarray}
\nonumber
\widehat{P}_m\Biggl\{ \frac 1{l_n^d}\sum_{i\in {\mathbb  L}_n}\Theta
_i({\mathbf  r} )\Lambda _{ij}\Biggr\} & = & \frac 1{l_m^d}\sum_{i\in
{\mathbb  L}_m}\sum_{i^{\prime }\in {\mathbb  L}_n}\Theta _i({\mathbf
r})\Lambda _{i^{\prime }j}\frac 1{l_n^d} \int\limits_{{\mathfrak
M}}d{\mathbf  r}^{\prime }\Theta _i({\mathbf  r}^{\prime })\Theta
_{i^{\prime }}({\mathbf  r}^{\prime })\\
\nonumber {} & = & \frac 1{l_m^d}\sum_{i\in
{\mathbb  L} _m}\Theta _i({\mathbf  r})\sum_{i^{\prime }\in {\mathbb
L}_n^i}\Lambda _{i^{\prime }j}=\frac 1{l_m^d}\sum_{i\in {\mathbb
L}_m}\Theta _i({\mathbf  r})\Lambda _{ij}.
\end{eqnarray}
Here we have taken into account that, first,

$$
\frac 1{l_n^d}\int\limits_{{\mathfrak  M}}d{\mathbf  r}^{\prime
}\Theta _i({\mathbf  r} ^{\prime })\Theta _{i^{\prime }}({\mathbf
r}^{\prime })=\left\{ \begin{array}{ccc} 1, & \mathrm{if} & i^{\prime
}\in {\mathbb  L}_n^i \\ 0, & \mathrm{if} & i^{\prime }\notin
{\mathbb  L}_n^i \end{array} \right. ,
$$
where ${\mathbb  L}_n^i$ is the full collection $\{i^{\prime }\}$ of
branches of level ${\mathbb  L}_n$ belonging to the subtree whose
stem is the branch $i$ and, thus, ${\mathfrak  M}_{i^{\prime
}}\subset {\mathfrak  M}_i$, and, second,

$$
\Lambda _{ij}=\sum_{i^{\prime }\in {\mathbb  L}_n^i}\Lambda _{i^{\prime }j}
$$
because of the conservation laws (\ref{a1.1}).
\qquad\end{proof}

In the present paper we consider the model for the supplying network
which envolves large but fininte number of hierarchy levels. In other
words, we regard the total number $N$ of levels as a fixed (but
large) value. In this case we can state the following.

\begin{proposition}
At leading order in the small parameter $\rho (n_i)/Z(n_i) $ for
almost all points ${\mathbf  r}\in {\mathfrak  M}$ it is fulfilled
the identity

\begin{equation}
\label{id}\sum\limits_{j\in {\mathfrak  N}^d}\Lambda
_{ij}R_{n_j}^0\Theta _j({\mathbf  r} )=\Theta _i({\mathbf  r}),
\end{equation}
provided the level number $n_i$ of the branch $i$ meets the
inequality $1\ll n_i\ll N$.
\end{proposition}

In this context the term ``leading order'' means, in addition, that
in (\ref {id}) we ignore any summand caused by variations of the
quantities $\rho (n)$ and $Z(n)$ when the argument $n$ changes on
scale of order unity. This is due to such summands cannot be analyzed
in the framework of the adopted continuous approximation of
Kirchhoff's equations.

Let us prove this assertion. \begin{proof} For a finite number of
hierarchy levels the volumentric measure of the interface made up of
the boundaries of all the fundamental domains is equal to zero. So in
proving (\ref{id}) we may ignore all the points belonging to this
interface. Therefore let ${\mathbf  r} $ be a point of the living
medium domain, ${\mathbf  r}\in {\mathfrak  M}$, that does not belong
to the boundary of any fundamental domain. For each level number $ n$
this point ${\mathbf  r}$ is located inside just one fundamental
domain of level $n$ which will be labelled as ${\mathfrak
M}_n({\mathbf  r})$. In this way we can build up the collection
$\{{\mathfrak  M}_n({\mathbf  r})\}$ ($n=0,1,\ldots ,N$) of all the
fundamental domains containing the point ${\mathbf  r}$
(Fig.~\ref{F10}).
These domains in turn specify the sequence $\{i_n({\mathbf
r})\}=\{i_o,i_1({\mathbf  r}),i_2({\mathbf  r} ),\ldots,i_N({\mathbf
r})\}$ of connected branches whose $n$-th term is the branch $
i_n({\mathbf  r})$ of level $n$ contained in the domain ${\mathfrak
M}_n({\mathbf  r})$. The branches $\{i_n({\mathbf  r})\}$ form a
continuous path ${\mathbb  P}({\mathbf  r})$ on the draining bed
$\mathfrak N^d$ that
leads from the stem $i_0$ to the branch $i_N({\mathbf  r})$ of the
last level.

\begin{figure}
\begin{center}
\epsfig{file=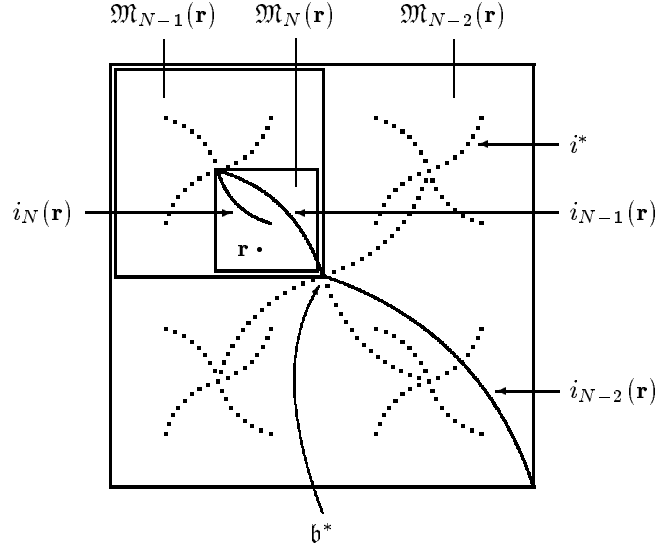}
\end{center}
\caption{Illustration of constructing the domain collection
$\{\mathfrak M_n(\mathbf r)\}$ and the collection of the corresponding
branches $\{i_n(\mathbf r)\}$ forming the path $\mathbb P(\mathbf r)$
$($solid line\/$)$ on the draining bed $($dotted lines\/$)$. $($For
simplicity is shown the two-dimensional living medium.$)$}
\label{F10}
\end{figure}

In formula (\ref{id}) the functions $\{\Theta _j({\mathbf  r})\}$
differ from zero for the branches $\{i_n({\mathbf  r})\}$ only, thus

\begin{equation}
\label{a2.2}\tilde \Lambda _i({\mathbf
r})\stackrel{\mathrm{def}}{\equiv } \sum\limits_{j\in {\mathfrak
N}_d}\Lambda _{ij}R_{n_j}^0\Theta _j({\mathbf  r} )=\sum\limits_{j\in
{\mathbb  P}({\mathbf  r})}\Lambda _{ij}R_{n_j}^0.
\end{equation}
Let us, first, consider a branch $i$ being one of the branches
$\{i_n({\mathbf  r} )\}$ whose level number $n_i$ meets the
inequality $N-n_i\gg 1$. For the given branch all the pairs
$\{i,i_n({\mathbf  r})\}$ belong to the first group specified in
Appendix~\ref{app:1}. Then taking into account expressions~(\ref
{*2.8}), (\ref{a1.5}), and (\ref{Gr1}) we can rewrite (\ref{a2.2}) as

\begin{eqnarray}\nonumber
\tilde \Lambda _i({\mathbf  r}) &=&
\sum_{j\in \mathbb P(\mathbf r)}^{n_j\geq n_i}\Lambda_{ij}R_{n_j}^0+
\sum_{j\in \mathbb P(\mathbf r)}^{n_j<n_i}\Lambda _{ij}R_{n_j}^0\\
\nonumber {}& = &
\frac 1{Z(n_i)}\sum\limits_{n_j=n_i}^N\rho
(n_j)+\sum\limits_{n_j=0}^{n_i-1}g^{-(n_i-n_j)}\frac{\rho (n_j)}{Z(n_j)}
\end{eqnarray}
and, thus,
\begin{equation}
\label{a2.3}\tilde \Lambda _i({\mathbf  r})=1+\sum\limits_{p=1}^{n_i}g^{-p}
\frac{\rho (n_i-p)}{Z(n_i-p)}\cong 1
\end{equation}
at leading order in the small parameter $\rho (n_i)/Z(n_i)$ for
$N-n_i\gg 1$.

Let us, now, consider a branch $i$ not belonging to the branch
collection $ \{i_n({\mathbf  r})\}$ (in Fig.~\ref{F10} it is the branch
$i^*$).  In this case there are a single node ${\mathfrak  b}$ (node
$\mathfrak b^*$ in Fig.~\ref{F10}) on the path ${\mathbb P}({\mathbf r})$ and a single
subtree that contains the branch $i$ and whose stem joins the path path
${\mathbb P}({\mathbf r})$ through the node $ {\mathfrak b}$.  The node
${\mathfrak b}$ devides the path ${\mathbb P}({\mathbf r})$ into two parts,
one going from the last level branch $i_N({\mathbf r})$ to the node
${\mathfrak b}$ and the other leading from the node ${\mathfrak  b}$ to the
stem of the draining bed. Let us ascribe to the node ${\mathfrak b}$ the
level number $n_{{\mathfrak b}}$ of branches going into it. Then we can say
that the former part of the path ${\mathbb P}({\mathbf  r})$ involves the
branches $\{i_n({\mathbf r})\}_{-}$ whose level numbers $n\geq
n_{{\mathfrak  b}}$.  For such branches the pairs $\{i,i_n({\mathbf
r})\}_{-}$ belong to the second group (see Appendix~\ref{app:1}) and $n_{
{\mathfrak b}}=n_{ii_n({\mathbf r})}$. The latter part is made up of the
branches $ \{i_n({\mathbf r})\}_{+}$ whose level numbers $n<n_{{\mathfrak
b}}$ and for these branches the pairs $\{i,i_n({\mathbf  r})\}_{+}$ belong
to the first group.  In the given case taking into account
expressions~(\ref{*2.8}), (\ref{a1.5}), (\ref{Gr1}), and (\ref{Gr2}) we can
rewrite (\ref{a2.2}) as

\begin{eqnarray}
\nonumber
\tilde \Lambda _i({\mathbf  r}) & = &
\sum_{j\in \mathbb P(\mathbf r)}^{n_j\geq n_{\mathfrak  b}}
\Lambda_{ij}R_{n_j}^0+
\sum_{j\in \mathbb P(\mathbf r)}^{n_j < n_{\mathfrak b}}
\Lambda _{ij}R_{n_j}^0 \\
\nonumber
& = & - \frac 1{(g-1)}g^{n_{{\mathfrak  b}}-n_i}\frac{\rho (n_{{\mathfrak
b}})}{[Z(n_{{\mathfrak  b}})]^2 }\sum_{n_j\geq n_{{\mathfrak
b}}}\rho (n_j)+\sum_{n_j<n_{{\mathfrak  b}}}g^{n_j-n_i}\frac{ \rho
(n_j)}{Z(n_j)} \\
\nonumber
& = & -\frac 1{(g-1)}g^{n_{{\mathfrak
b}}-n_i}\frac{\rho (n_{ {\mathfrak  b}})}{Z(n_{{\mathfrak
b}})}+g^{n_{{\mathfrak  b}}-n_i}\sum_{p=1}^{n_{{\mathfrak  b}}}g^{-p}
\frac{\rho (n_{{\mathfrak  b}}-p)}{Z(n_{{\mathfrak  b}}-p)}.
\end{eqnarray}
Whence treating the small ratio $\rho (n_{{\mathfrak
b}})/Z(n_{{\mathfrak  b}})$ as a smooth function of the continuous
variable $n$ we get for $n_{{\mathfrak  b}}\gg 1$

\begin{equation}
\label{a2.4}\tilde \Lambda _i({\mathbf  r})\simeq -g^{n_{{\mathfrak
b}}-n_i}\frac d{dn }\left[ \frac{\rho (n)}{Z(n)}\right]
_{n=n_{{\mathfrak  b}}}\sum\limits_{p=1}^ \infty pg^{-p}
\end{equation}
at leading order in this small parameter.  Therefore we may state
that $ \tilde \Lambda _i({\mathbf  r})=o(\frac{\rho
(n_i)}{Z(n_i)})$ because in the given case $n_{{\mathfrak  b}}<n_i$
and so either $n_{{\mathfrak  b}}\sim n_i$ or
$n_i - n_{{\mathfrak b}}\gg 1$
and for such values of $n_{{\mathfrak  b}}$ the magnitude of
$\tilde \Lambda _i({\mathbf  r})$ is of exponential smallness:
$\tilde \Lambda _i({\mathbf  r })\sim g^{n_{{\mathfrak  b}}-n_i}$.
The given nonzero value of $\tilde \Lambda _i({\mathbf r})$ is actually
due to variations of the function $\rho (n)/Z(n)$ as the argument $n$
changes on scales of order unity, so in this case we have to set
$\tilde \Lambda _i({\mathbf  r})=0$.

Summarizing the aforesaid we see that if the point ${\mathbf
r}$ is located inside the fundamental domain ${\mathfrak  M}_i$
corresponding to the branch $i$, i.e. $\Theta _i({\mathbf  r})=1$
then the branch $i$ belongs to the path ${\mathbb  P}({\mathbf  r})$
and the value $\tilde \Lambda _i({\mathbf  r})=1$ at leading order in
$\rho (n)/Z(n)$. Otherwise, when ${\mathbf  r}\notin \mathfrak  M
_i$ and so $i\notin {\mathbb  P}({\mathbf  r})$ the value $\Theta
_i({\mathbf  r})=0$ and we have to set $\tilde \Lambda _i({\mathbf
r})=0$. This substantiates identity~(\ref{id}).\qquad\end{proof}

%\newpage

\end{document}